\documentclass[review]{elsarticle}
\usepackage{graphicx}
\usepackage{amsmath}
\usepackage{amssymb}
\usepackage{bm}
\usepackage{xcolor}

\usepackage{ulem} 

\DeclareMathOperator\dv{div} 
\DeclareMathOperator\id{I}   
\DeclareMathOperator\tr{tr}   
\newcommand{\bu}{\mathbf{u}}  
\newcommand{\bq}{\mathbf{q}}  
\newcommand{\bx}{\mathbf{x}}  
\newcommand{\be}{\mathbf{e}}  
\newcommand{\bbs}{\mathbf{s}}  
\newcommand{\bn}{\mathbf{n}}  

\newcommand{\bpsi}{\bm{\psi}} 
\newcommand{\bsigma}{\bm{\sigma}} 
\newcommand{\btau}{\bm{\tau}} 
\newcommand{\epss}{{\cal E}}

\newcommand{\pd}[2]{ 
	\dfrac{\partial #1}{\partial #2}
} 

\newcommand{\pddB}[2]{ 
	\dfrac{\partial }{\partial #2}\Bigl(#1\Bigr)
} 
 
\graphicspath{{img/}}

\begin{document}	
	
    \begin{frontmatter}
        
        \title{Influence of pore pressure to the development of a hydraulic fracture in poroelastic medium}
       
        \author[address_NSU,address_hydro]{S.V. Golovin\corref{cor1}}
        \cortext[cor1]{Corresponding author at: Lavrentyrev Institute of Hydrodynamics, pr. Lavrentyeva 15, Novosibirsk 630090, Russia. Fax: +7 (383) 333 1612}
        \ead{golovin@hydro.nsc.ru}
             
        \author[address_NSU]{A.N. Baykin}
        
        \address[address_hydro]{Lavrentyrev Institute of Hydrodynamics, pr. Lavrentyeva 15, Novosibirsk 630090, Russia}
        \address[address_NSU]{Novosibirsk State University, ul. Pirogova 2, Novosibirsk 630090, Russia}

    	\begin{abstract}
            In this paper we demonstrate the influence of the pore pressure to the development of a hydraulically-driven fracture in a poroelastic medium. We present a novel numerical model for propagation of a planar hydraulic fracture and prove its correctness by demonstration of the numerical convergence and by comparison with known solutions. The advantage of the algorithm is that it does not require the distinguishing of the fracture's tips and reconstruction of the numerical mesh according to the fracture propagation. Next, we perform a thorough analysis of the interplay of fluid filtration and redistribution of stresses near the fracture. We demonstrate that the fracture length decreases with the increase of the Biot's number (the parameter that determines the contribution of the pore pressure to the stress) and explain this effect by analysing the near-fracture pore pressure, rock deformation and stresses. We conclude, that the correct account for the fluid exchange between the fracture and the rock should be based not only on physical parameters of the rock and fluid, but also on the analysis of stresses near the fracture.
        \end{abstract}
        
        \begin{keyword}
            Hydraulic fracture, Poroelasticity, Pore pressure influence, Finite element method
        \end{keyword}
        
    \end{frontmatter}


\section{Introduction}
Mathematical modelling of hydraulically-driven fractures is a highly demanded subject in modern technologies for enhancement of reservoir permeability in hydrocarbon production as well as in geophysical problems related, for instance, to the development of magmatic dykes. Recent progress in the modelling of hydraulic fracture dynamics is described in the review papers \cite{Adachi_Siebrits_Pierce_Desroches_2007, Detournay_2016} and citations therein. The early although widely used models by Khristianovich, Zheltov, Geertsma, and de Klerk (KGD) \cite{Z-K, GeertsmadeKlerk1969} and by  Perkins, Kern and Nordgen (PKN) \cite{PerkinsKern1961, Nordgren1972} assume that the fracture is propagating in infinite elastic medium and the fluid exchange between the fracture and the porous reservoir is modelled as only a fracturing fluid loss (leakoff) according to Carter's formula \cite{Economides} which proposes that the leakoff is inverse proportional to the square root of the wetting time. More advanced models of the leakoff suppose computation of the pore pressure around the fracture by solving the piezoconduction equation \cite{Gordeyev_Entov_1997} although still do not considering the influence of the pore fluid to stresses. 

Theoretical study of the action of the pore pressure to the distribution of stresses near the fracture was carried out in many papers, a detailed review can be found in the Introduction of the dissertation by Y. Yuan \cite{Yuan_PhD_1997}. In particular, the additional stiffness of the rock due to the pressure in the vicinity of the fracture was treated as the backstress  \cite{Vandamme_Roegiers, Kovalyshen_PhD}. It was noted, that the wellbore fluid pressure needed to open the fracture considerably rises due to the backstress. The same effect leads to the overestimation of the minifrac tests for the {\it in situ} minimal principal stress \cite{Boone_Ingraffea_Roegiers_1991, Detournay_Cheng_Roegiers_McLennan_1989}. The mentioned facts indicate that proper account for the action of the pore pressure and proper modelling of the fluid exchange between the fracture and the porous reservoir is principal for the correct description of the fracture dynamics. 

In our paper we propose a mathematical model for propagation of a hydraulic fracture in a poroelastic medium. The numerical solution of the problem is carried out by the finite element method with the use of a modification of the algorithm suggested in \cite{ShelBaikGol2014}. We use an approach of modelling free of explicit tracking of the fracture's tip similar to the one used in \cite{GolovinIsaevEtal2015}. The advantage of our model is that we do not need to rebuild the computational mesh according to propagation of the fracture, which is typical for problems of this type. The rock failure criteria is modelled using the cohesive zone model initially proposed by Barenblatt \cite{Barenblatt} and Dugdale \cite{Dugdale}. This model allows us to eliminate the stress singularity at the fracture's tip as well as to integrate the computation of the failure criteria into the numerical algorithm. The correctness of the model is checked by the analysis of the numerical convergence of the algorithm and by comparison with analytic and numerical solutions presented in \cite{Carrier_Granet}. In all observed cases we have a satisfactory coincidence of the solutions.

The constructed model is used for the analysis of the influence of the pore pressure to the fracture dynamics. We demonstrate that the dynamics is governed by the two factors: the rate of the medium displacement that modifies the filtration, and by the backstress that significantly increases the pressure inside the fracture. For the relatively high rock permeability these two factors notably increase the leakoff and hence, decrease the length of the fracture. The demonstrated effect is dumped by high reservoir's storage coefficient or low rock permeability.

\section{Mathematical formulation of the problem}\label{sec:problem_formul}
Let us consider a vertical planar fracture of fixed height $H$, propagating along the straight line denoted as $x$-axis (see Figure~\ref{fig:frac_geometry_3d}). We direct $z$-axis upwards and $y$-axis perpendicular to the plane of the fracture propagation. We suppose that fracture's aperture is constant along the vertical coordinate $z$, so the plain strain approximation is applicable. This implies, that we can limit ourselves to observing only the central cross-section $z=0$ of the fracture.
		
\subsection{Equations for the poroelastic reservoir}\label{ReservoirEqs}
The poroelastic medium is characterized by its porosity $\phi$ and permeability $k_r(\bx)$, with the solid phase displacement $\bu(t,\bx)$, and the pore pressure $p(t,\bx)$. Pores are saturated by a single-phase Newtonian fluid with the effective viscosity $\eta_r$. We make use of the linear Darcy law for the fluid velocity $\bq=-(k_r/\eta_r)\nabla p$. It is supposed that the fluid filtrating from the fracture to the reservoir has the same viscosity as the pore fluid. However, the fluid within the fracture has different viscosity $\eta_f$.  This corresponds to the normal situation in hydraulic fracturing when the fracturing fluid is a high-viscous gel and only its low-viscous base fluid is filtrated into the reservoir.
	
For the generality,  the reservoir is initially subjected to a prestress with the stress tensor $\btau^0(x,y)$. Since we observe only straight fractures, tensor $\btau^0$ satisfies symmetry conditions relative to $x$-axis.      
	
The governing equations of the quasi-static poroelasticity model are the following \cite{Coussy}:
	\begin{equation}\label{eq:model}
        \begin{array}{l}
            \displaystyle \dv{\btau}=0, \quad \btau = \btau^0 + \lambda \dv{\bu} \id+2\mu\, \epss{(\bu)}-\alpha p \id,\\[4mm]
            \displaystyle S_\varepsilon\pd{p}{t} = \dv{\Bigl( \frac{k_r}{\eta_r}\nabla p - \alpha \pd{\bu}{t} \Bigr)}.
        \end{array}
    \end{equation}
Here $\epss(\bu)$ is the Cauchy's strain tensor
$2\epss(\bu)_{ij}=\partial u_i/\partial x_j+\partial u_j/\partial x_i$ $(i,j=1,2)$,
$\alpha$ is the Biot coefficient,
$\lambda(\bx)$ and $\mu(\bx)$ are  elasticity moduli, $\id$ is the identity tensor. The storativity $S_\varepsilon$ reflects the dependence of the Lagrangian porosity  $\phi$ on $\epsilon = \tr\epss$ and $p$ as in~\cite{Coussy}:
	\begin{equation}\label{eq:storativity_def}    
	\pd{\phi}{t}=\alpha \pd{\epsilon}{t}+S_\varepsilon\pd{p}{t},\quad 		S_\varepsilon = \dfrac{(\phi_0-\alpha)(1-\alpha)}{K},
	\end{equation} 
where $K = \lambda + \dfrac{2\mu}{3}$ is the bulk modulus, $\phi_0$ is the initial porosity. Due to the plane strain approximation, the solid phase displacement vector $\bu = (u_1, u_2) = (u, v)$
is two-dimensional, all vector operations are also taken in 2D space of independent variables $\bx=(x_1, x_2) = (x, y)$.
	
\begin{figure}[h]
		\begin{minipage}[b]{0.45\linewidth}	
		\centering
		\includegraphics[width=\linewidth]{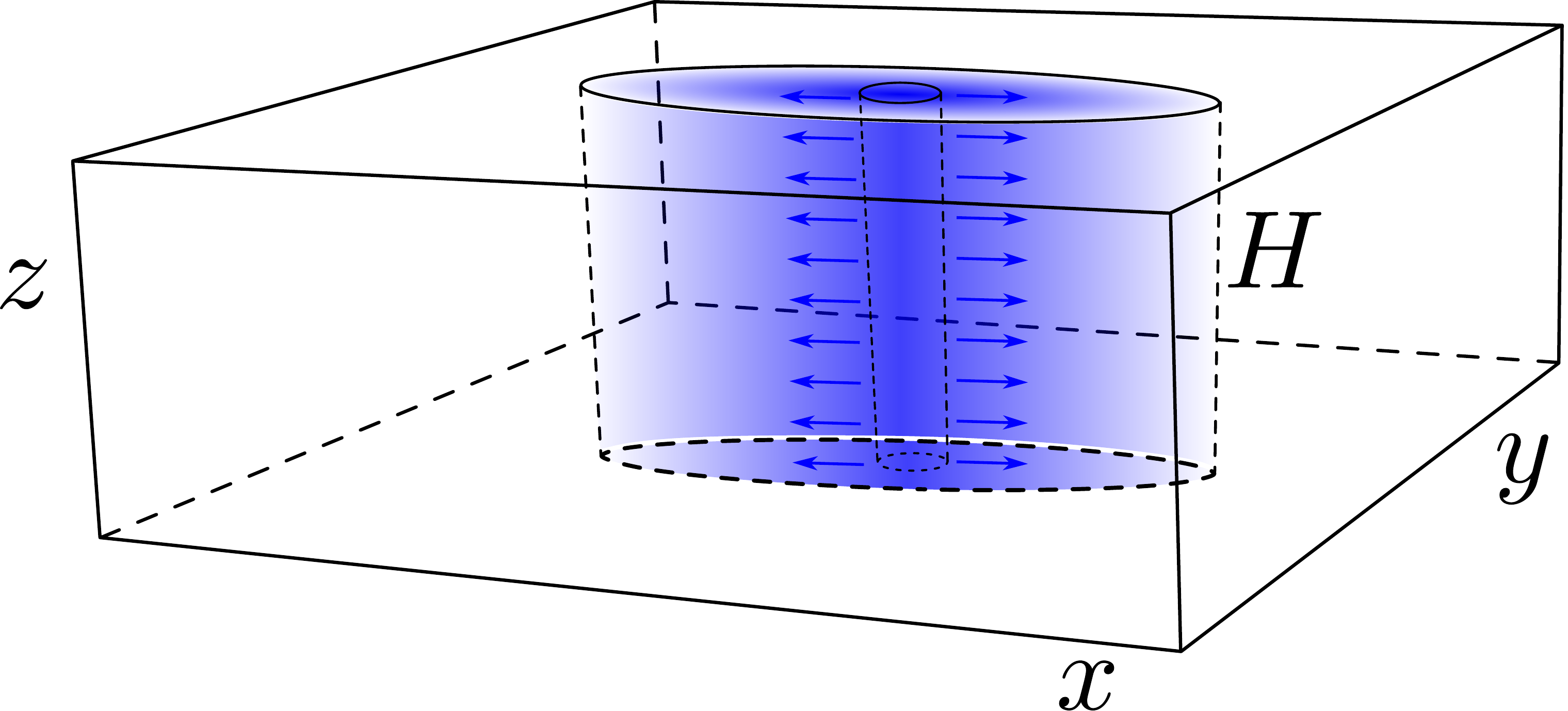}
		\caption{Planar vertical hydraulic fracture in a poroelastic medium}
		\label{fig:frac_geometry_3d}
		\end{minipage}
		\hfill
		\begin{minipage}[b][][b]{0.45\linewidth}
			\centering
			\includegraphics[width=0.95\linewidth]{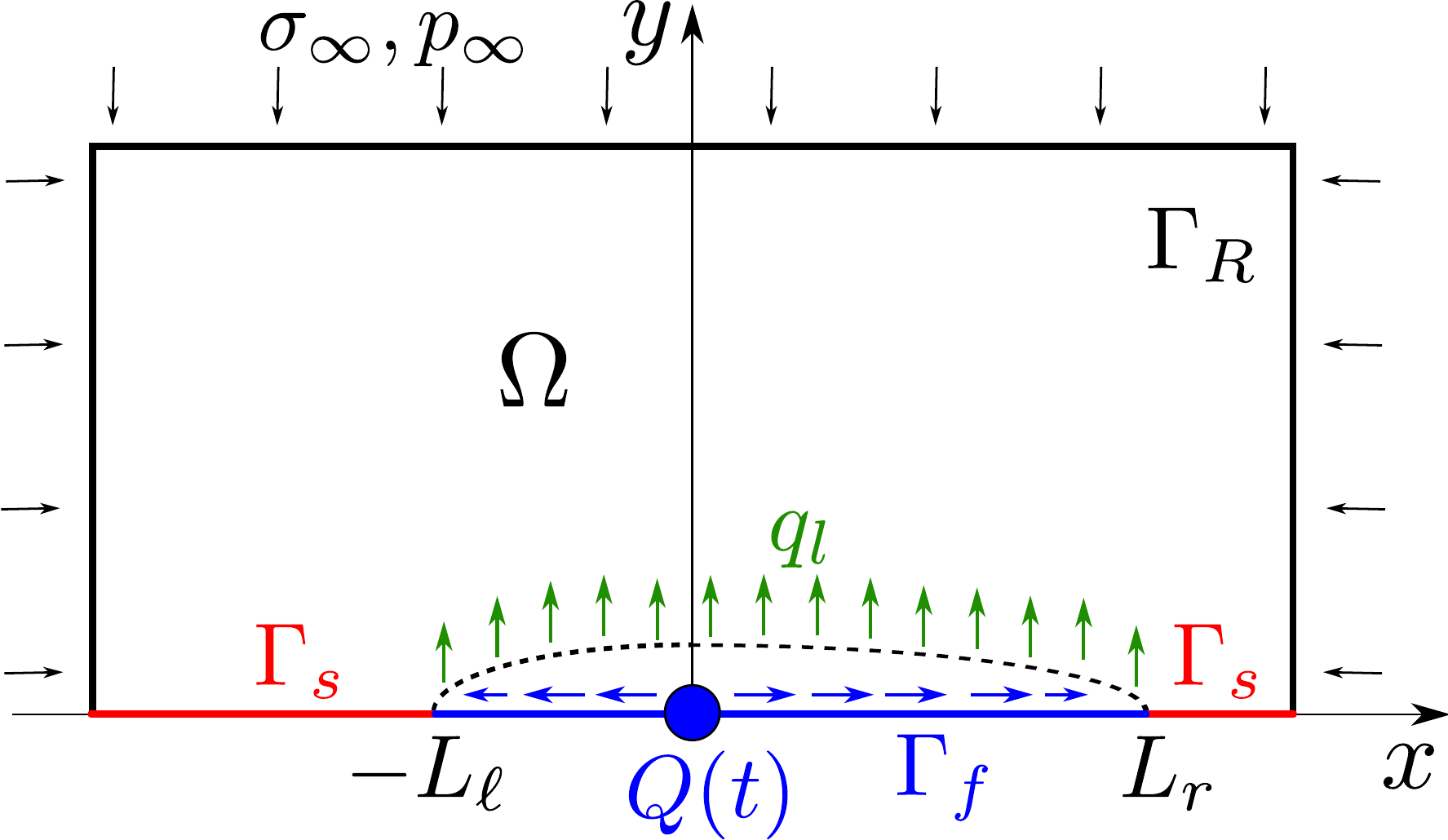}
			\caption{The horizontal  cross-section of the fracture by plane $z=0$}
			\label{fig:fracture_geometry_2d}
		\end{minipage}
\end{figure}

Symmetry of the problem with respect to $Ox$-axis allows solving equations \eqref{eq:model} in domain $\Omega = \{(x,y): |x|\leq R, 0\leq y\leq R\}$ as shown in Figure \ref{fig:fracture_geometry_2d}.
Over the outer boundary $\Gamma_R=\{\partial\Omega\cap y>0\}$ the 
confining far-field stress  $\bsigma_\infty$ is applied and the  constant pore pressure $p=p_\infty$ is prescribed:
\begin{equation}\label{eq:GammaRBC}
	    \Gamma_R:\quad p=p_\infty,\quad \btau\langle\bn\rangle=\bsigma_\infty,
	    \quad (\btau\langle\bn\rangle)_i=\tau_{ij}n_j.
\end{equation}

Henceforth $\bn$ and $\bbs$ denotes the outer normal and tangential unit  vectors to the boundary of the domain $\Omega$; the summation over the repeating indices is implied. We restrict ourselves to the case  $\bsigma_\infty = -\sigma_\infty \be_2$, where $\sigma_\infty$ is the minimal principal {\it in situ} stress. Moreover, we assume that the prestress $\btau^0$ satisfy the same boundary condition: $\btau^0\langle\bn\rangle|_{\Gamma_R} = \bsigma_\infty$.

In order to close the model it is supplemented with the initial data at some moment $t^0$:
\begin{equation}\label{eq:init_cond}
\bu|_{t=t^0} = \bu^0(x,y), \quad p|_{t=t^0} = p^0(x,y),\quad L_i|_{t=t^0} = L^0_i,\; i=\ell,r.
\end{equation}

\subsection{Equations for the hydraulic fracture}
 
The line $y=0$ is divided into the part $\Gamma_f = \{ -L_\ell(t) \leqslant x \leqslant L_r(t),\; y=0 \}$ occupied by the fracture, and the remaining part $\Gamma_s = \{ -R<x<-L_\ell(t),\; y=0\}\bigcup \{ L_r(t)<x<R,\; y=0\}$. Outside the fracture on  $\Gamma_s$ the symmetry conditions (see~\cite{ShelBaikGol2014}) are satisfied:
\begin{equation}\label{eq:symmBC}
		\Gamma_s:\quad \pd{u}{y}=0, \quad v=0, \quad \pd{p}{y}=0.
\end{equation}

With $p_f(t,x)$ standing for the fluid pressure inside the fracture and $\sigma_{coh}$ denoting the cohesive forces near the fracture's tips (explained below), the force balance over the fracture’s wall yields
	\begin{equation}\label{eq:fracBC}
		\Gamma_f:\quad p=p_f, \quad \bn\cdot\btau\langle\bn\rangle=-p_f + \sigma_{coh}, \quad \bbs\cdot\btau\langle\bn\rangle=0.
	\end{equation}
Here we neglect the tangential stress due to the fluid friction on the	fracture’s walls in comparison with the normal stress.

	The fluid flow in the fracture is governed by the mass conservation
	law complemented with the Poiseuille formula:	
	\begin{equation}\label{eq:lubric_equation}
		\pd{w}{t}+\pd{(wq)}{x}=-q_l,\quad w\equiv v|_{y=0},\quad q=-\frac{(2w)^2}{12\eta_f}\pd{p_f}{x}.
	\end{equation}
	Here $w$ is a half of the fracture aperture, $q$ is the fluid velocity in $x$-direction. No fluid lag is assumed at the fracture's tip.
	
	The leakoff velocity $q_l$ is given by the Darcy law as
	\begin{equation}\label{leakoff}
		q_l=-\left.\frac{k_r}{\eta_r}\pd{p}{y}\right|_{y=0}.
	\end{equation}
	
	The resulting equation governing the flow inside the fracture reads
	\begin{equation}\label{eq:mass_conserve_lubricBC}
		\pd{w}{t}=\pd{}{x}\left(\frac{w^3}{3\eta_f}\pd{p_f}{x}\right)+\left.\frac{k_r}{\eta_r}\pd{p}{y}\right|_{y=0}.
	\end{equation}
	
	The flow rate (per unit height) injected into the fracture upper half-plane is calculated as 
	\begin{equation}\label{eq:flowrateBC}
		Q(t) = \frac{Q_v(t)}{2H}= -\left.\frac{w^3}{3\eta_f}\pd{p_f}{x}\right|_{x=0+} + \left.\frac{w^3}{3\eta_f}\pd{p_f}{x}\right|_{x=0-},
	\end{equation}
	where the division by 2 shows that the total flow rate is equally distributed between the symmetric fracture parts, and $Q_v(t)$ denotes the volumetric flow rate injected into the well.

Equation \eqref{eq:mass_conserve_lubricBC} is often referred to as the lubrication theory equation~\cite{Adachi_Siebrits_Pierce_Desroches_2007}. Note that, due to the right-hand side of \eqref{leakoff}, equation \eqref{eq:mass_conserve_lubricBC} represents a boundary condition for equations of the main model \eqref{eq:model}. The leakoff rate $q_l$ arises here naturally in the course of the problem’s solution, which differentiates the model favourably from the usual artificial approximations like Carter’s formula or other similar expressions~\cite{Economides}.

\subsection{The failure criteria}

\begin{figure}[h]
	\begin{minipage}[t]{0.5\linewidth} 
		\centering
		\includegraphics[width=0.8\linewidth]{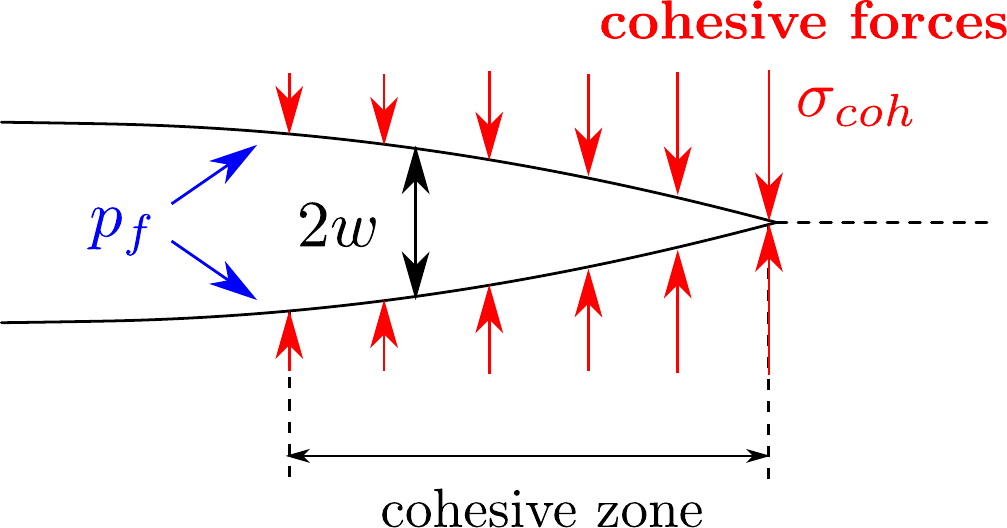}
		\caption{ Cohesive zone near the fracture's tip }
		\label{fig:cohesive_zone_sketch}
	\end{minipage}
	\hspace{0.5cm}
	\begin{minipage}[t]{0.5\linewidth}
		\centering
		\includegraphics[width=0.8\linewidth]{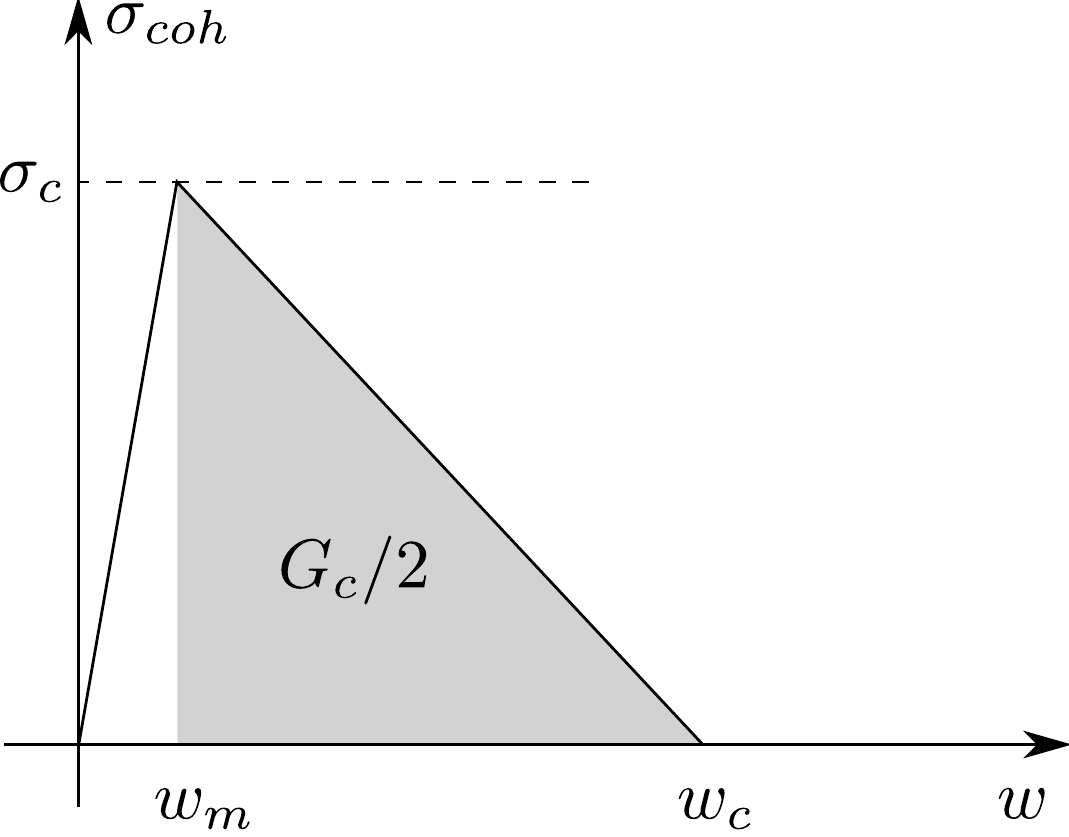}
		\caption{ Bi-linear traction/separation law }
		\label{fig:cohesive_law}
	\end{minipage}
\end{figure}

In order to account for the rock toughness during the fracturing, we adopt the  cohesive zone model initially proposed by Barenblatt \cite{Barenblatt} and Dugdale \cite{Dugdale}. In this approach it is postulated the existence of cohesive forces $\sigma_{coh}$ (see Figure \ref{fig:cohesive_zone_sketch}) acting in the zone of micro-cracking and plastic deformations in the vicinity of the fracture's tip. The cohesive forces slow down the fracture opening process at the initial stage of the rock failure. On the computational side, presence of the cohesive forces removes the stress singularity at the fracture's tip inherent to the Linear Elastic Fracture Mechanics (LEFM) by making the fracture aperture smoothly vanishing towards the tip.  

We use the following traction/separation bi-linear law \cite{Geubelle} to reflect the dependence of $\sigma_{coh}$ on the fracture aperture $2w$ as shown in Figure \ref{fig:cohesive_law}:
\begin{equation}\label{eq:cohesive_law}
\sigma_{coh}(w)=\left\{
\begin{array}{ll}
\sigma_c\dfrac{w}{w_m}, & 0\leqslant w\leqslant w_m, \\[2ex]
\sigma_c\Bigl(\dfrac{w_c-w}{w_c-w_m}\Bigr), & w_m\leqslant w \leqslant w_c,\\[2ex]
0, & w \geqslant w_c.\\[2ex]
\end{array}
\right.
\end{equation}

The cohesive forces reach their maximum value $\sigma_c$ near the fracture's tip as the fracture's aperture is equal to $w_m$. The region of the micro-cracking is limited by maximal fracture's aperture $w_c$. The value of $w_c$ is calculated from the considerations that the energy released during the creation of the new fracture surface is equal to the work of the cohesive forces on the fracture opening. Hence,
\begin{equation}\label{eq:cohesive_energy}
G_c = \sigma_c\,w_c, 
\end{equation} 	
where $G_c$ is the fracture energy in the Griffiths's theory of brittle fractures \cite{Griffith}.
The elastic region of cohesive forces is small, $w_m = 5\times10^{-4}w_c$. It is required to regularize the cohesive energy near $w=0$ \cite{Carrier_Granet}.

If the cohesive zone is small relative to the fracture's length, the connection with the fracture toughness $K_{Ic}$ from LEFM is given by Irvin's formula \cite{Irwin}:
\begin{equation}\label{eq:irvin_formula}
K_{Ic} = \sqrt{G_c\dfrac{E}{1-\nu^2}}, 
\end{equation} 	
where $E$ is the Young's modulus and $\nu$ is the Poisson's ratio.

\subsection{The full set of equations}\label{FullSetEqs}
	
For computational reasons, it is convenient to homogenise the conditions over the outer boundary $\Gamma_R$. It can be done by considering the stresses inside the reservoir relative to the prestress state $\btau^0$, and taking  $p_\infty$ as a reference pressure. For the boundary conditions defined in Section \ref{ReservoirEqs} the initial deformation $\bu^0=(u^0,v^0)$ due to the prestress has the form
\[u^0=\frac{\lambda \sigma_\infty +2 \alpha  \mu  p_\infty}{4 \mu(\lambda+ \mu)}\, x,\quad v^0=\frac{-\sigma_\infty  (\lambda +2 \mu )+2 \alpha  \mu  p_\infty}{4 \mu  (\lambda +\mu )}\, y.\]
At that, 
\[\btau^0 = \left(
\begin{array}{cc}
0&0\\
0&-\sigma_\infty
\end{array}\right)\]
Similar to \cite{ShelBaikGol2014}, the following new sought functions are introduced:
	\begin{equation}\label{eq:tau_tild}
		\tilde{\bu} = \bu -  \bu^0, \quad \tilde{\btau} = \btau -\btau^0, \quad \tilde{p} = p - p_\infty.
	\end{equation}	
	Substituting \eqref{eq:tau_tild} into equations \eqref{eq:model} and taking into account boundary conditions \eqref{eq:GammaRBC}, \eqref{eq:symmBC}, \eqref{eq:fracBC},  \eqref{eq:mass_conserve_lubricBC}, \eqref{eq:flowrateBC}, we obtain the following problem
	\begin{flalign}
		\displaystyle \Omega: & \quad \dv{\tilde{\btau}}=-\dv\btau^0, \quad \tilde{\btau} =  \lambda \dv{\tilde{\bu}}\id+2 \,\mu\, \epss{(\tilde{\bu})}-\alpha \tilde{p} \id, & \label{eq:tild_formul_equilib} \\[4mm]
		\displaystyle \Omega: & \quad S_\varepsilon\pd{\tilde{p}}{t} =  \dv{\Bigl( \dfrac{k_r}{\eta_r} \nabla \tilde{p} - \alpha \pd{\tilde{\bu}}{t} \Bigr)},& \label{eq:tild_formul_filtr}  \\[4mm]
		\displaystyle \Gamma_R: & \quad \tilde{p}=0, \quad  \tilde{\btau}\langle\bn\rangle = 0, & \label{eq:tild_formul_gamma_R}  \\[4mm]
		\displaystyle \Gamma_s: & \quad \pd{\tilde{u}}{y}=0, \quad \tilde{v} = 0, \quad \pd{\tilde{p}}{y} = 0, & \label{eq:tild_formul_gamma_s}  \\[4mm]
		\displaystyle \Gamma_f: & \quad  \bn \cdot \tilde{\btau}\langle\bn\rangle = -(\tilde{p}+p_\infty) - \bn \cdot \btau^0\langle\bn\rangle + \sigma_{coh} ,\quad \bbs \cdot \tilde{\btau}\langle\bn\rangle = 0, & \label{eq:tild_formul_gamma_f_force} \\[4mm]
		\displaystyle \Gamma_f: & \quad \pd{\tilde{v}}{t}=\pd{}{x}\left(\frac{\tilde{v}^3}{3\eta_f}\pd{\tilde{p}}{x}\right)+\frac{k_r}{\eta_r}\pd{\tilde{p}}{y}; \quad -\frac{\tilde{v}^3}{3\eta_f}\pd{\tilde{p}}{x}\Bigr|_{y=0,x=0+} + \frac{\tilde{v}^3}{3\eta_f}\pd{\tilde{p}}{x}\Bigr|_{y=0,x=0-} 
		= Q(t). & \label{eq:tild_formul_gamma_f_lubric} 		
	\end{flalign}
	
	The initial data at $t=t^0$ is the following:
	\[
	\tilde{\bu}|_{t=t^0} = 0, \quad \tilde{p}|_{t=t^0} = 0,\quad L_i|_{t=t^0} = L^0_i,\; i=\ell,r.
	\]
	In the remaining part of the paper we work with the new sought functions skipping
	the tilde for simplicity of notations.

\section{Numerical algorithm}
	In this Section we provide the numerical method to solve the problem stated in Section \ref{sec:problem_formul}. We start form the weak formulation of the problem. Following \cite{ShelBaikGol2014}, let us choose a smooth vector-function $\bpsi=\bigl(\psi_1(x,y),\psi_2(x,y)\bigr)$  and a smooth scalar function $\varphi(x,y)$ such that	
	\begin{equation}
		\psi_2|_{\Gamma_s} = 0,\quad  \varphi|_{\Gamma_R} = 0.
	\end{equation}
	Then we multiply equation \eqref{eq:tild_formul_equilib}
	and  \eqref{eq:tild_formul_filtr} by $\bpsi$ and $\varphi$ respectively and integrate over $\Omega$. 
	Taking into account the boundary conditions
	 \eqref{eq:tild_formul_gamma_R}--\eqref{eq:tild_formul_gamma_f_lubric} after integration we  obtain

	\begin{multline}\label{eq:variat_formul_equil}
	      \displaystyle \int\limits_{\Omega}\bigl(\lambda\dv{\bigl(\bu\bigr)}-\alpha\,p\bigr)\dv{\bigl(\bpsi\bigr)}+2 \mu\,\mathcal{E}\bigl(\bu\bigr):\mathcal{E}\bigl(\bpsi\,\bigr)\,dx dy\, -  \\[1ex]
	       \displaystyle - \int\limits_{\Gamma_f}(p+p_\infty + \bn \cdot \btau^0\langle\bn\rangle - \sigma_{coh})\psi_2\,dx = 0,	    
	\end{multline}  
	\begin{multline}\label{eq:variat_formul_filtr}
	\displaystyle  \int\limits_{\Omega}S_\varepsilon \pd{p}{t}\varphi\,dx dy +\int\limits_{\Omega}\dfrac{k_r}{\eta_r}\nabla p\cdot\nabla \varphi\,dx dy +\int\limits_{\Omega}\alpha\,\pddB{\dv{\bu}}{t} \varphi\,dx dy + \\[2ex]
	 \displaystyle  +\int\limits_{\Gamma_f} \pd{v}{t}  \varphi\, dx+
	 \int\limits_{\Gamma_f} \dfrac{v^3}{3\eta_f}\,p_{x}\varphi_{x} \, dx- Q(t) \varphi(0,0)=0.
	\end{multline}  
This formulation is not convenient for the computational use because of the necessity to track the fracture's tips and change the boundary $\Gamma_f$ at every time step when the fracture changes its size. 

In order to fix the computational domain, we use the method similar to the one proposed in \cite{GolovinIsaevEtal2015}. Namely, we treat the set  $\Gamma_f$ as a path of the potential fracture propagation that has a closed part where $v=0$ and an opened part where $v>0$. However, under such interpretation we cannot guarantee the absence of the interpenetration of the opposite fracture walls during the computations. In order to avoid this problem we impose an additional restriction to the problem, formulated in Section  \ref{FullSetEqs}:
\begin{equation}\label{eq:restr}
\Gamma_f:\quad v\geqslant 0.
\end{equation}
In order to make use of the restriction \eqref{eq:restr} we add a penalty term
\begin{equation}\label{eq:penalty_term}
\dfrac{1}{\delta}\int\limits_{\Gamma_f}\chi_{[v<0]}\,v\,\psi_2\,dx
\end{equation}
to the weak formulation \eqref{eq:variat_formul_equil}. Here $\delta\ll 1$ is a small number and $\chi_{[v<0]}$ is the indicator function of the set~$\{\bx:v(\bx)<0\}$. Equation \eqref{eq:variat_formul_equil} is transformed into the following one
	\begin{multline}\label{eq:variat_formul_equil_penalty}
		\displaystyle \int\limits_{\Omega}\bigl(\lambda\dv{\bigl(\bu\bigr)}-\alpha\,p\bigr)\dv{\bigl(\bpsi\bigr)}+2 \mu\,\mathcal{E}\bigl(\bu\bigr):\mathcal{E}\bigl(\bpsi\,\bigr)\,dx dy\, -  \\[1ex]
		\displaystyle - \int\limits_{\Gamma_f}(p+p_\infty + \bn \cdot \btau^0\langle\bn\rangle - \sigma_{coh})\psi_2\,dx	+\dfrac{1}{\delta}\int\limits_{\Gamma_f}\chi_{[v<0]}\,v\,\psi_2\,dx = 0.    
	\end{multline} 	
	Introduction of the penalty term is equivalent to the replacement of the boundary condition \eqref{eq:tild_formul_gamma_f_force} to the condition
	\begin{equation}\label{eq:variat_formul_gamma_f_force_penalty}
		\displaystyle \Gamma_f:  \quad  \bn \cdot \btau\langle\bn\rangle = -(p+p_\infty)+\dfrac{1}{\delta}\chi_{[v<0]}\,v - \bn \cdot \btau^0\langle\bn\rangle+\sigma_{coh} ,\quad \bbs \cdot \btau\langle\bn\rangle = 0. 
	\end{equation}
	Equation \eqref{eq:variat_formul_gamma_f_force_penalty} at $\delta\to 0$ is satisfied only for non-negative values of $v$ over $\Gamma_f$.
	The locations of the fracture's tips can be calculated as $L_l(t) = \inf\limits_{v|_{y=0}> 0} x $ and $L_r(t) = \sup\limits_{v|_{y=0}>0} x$.  Integrals over $\Gamma_f$ in equation \eqref{eq:variat_formul_filtr},  are non-zero only over $\Gamma_f\cap\{\bx:v(\bx)>0\}$, hence, the lubrication equation is solved only over the opened part of the fracture; over the closed part with $v=0$ the lubrication equation and the corresponding integrals over $\Gamma_f$ in equation \eqref{eq:variat_formul_filtr} vanish identically.
	
	The problem \eqref{eq:variat_formul_filtr}, \eqref{eq:variat_formul_equil_penalty} is  solved using finite element method via the open-source FEM package FreeFEM++ \cite{FreeFEM}. For spacial discretization we use the piecewise-linear $P1$ elements over the triangulated computational domain $\Omega$. The time derivatives are approximated with the first order of accuracy: $\pd{f}{t} \approx \dfrac{f^{n+1}-f^n}{\Delta t}$, where $f$ denotes either of the functions  $u$, $v$ or $p$; $\Delta t$ is a time step. The upper index designates the number of the time step: $f^n = f(t^n,\bx)$, $t^n = t^0 + n\Delta t$. The computations start with the intact state $\bu^0 = 0$ and $p^0=0$  as the initial data. The non-linearity is resolved by the Newton-Raphson method.

\section{Verification of the numerical algorithm }

\subsection{Numerical convergence test}
		\begin{table}[!t]
		\renewcommand{\arraystretch}{1.2} 
			\begin{center}	
				\caption {Input parameters for the reference verification case} \label{tab:sim_parameters_verification}				\begin{tabular}{|@{\hspace{2em}}l@{\hspace{2em}}|@{\hspace{2em}}c@{\hspace{2em}}|}
					
					\hline	
					{\bf Parameter} & {\bf Value} \\
					\cline{1-2}
					Domain size, $R$  &  $105$ m\\
					Max. right tip position, $L_r^{max}$  &  $40$ m\\
					Max. left tip position, $L_\ell^{max}$  &  $40$ m\\
					Young's modulus, $E$  &  $17$ GPa\\
					Poisson's ratio, $\nu$ &  $0.2$ \\
					Fracture energy, $G_c$ & 120 Pa$\cdot$m \\
					Critical cohesive stress, $\sigma_c$ & 1.25 MPa \\
					Initial porosity, $\phi$ & 0.2 \\
					Reservoir permeability, $k_r$ &  $10^{-14}$ m$^2$\\ 
					Biot coefficient, $\alpha$ &  $0.75$ \\
					Storage coefficient, $S_\varepsilon$ &  $1.46\times 10^{-11}$ Pa$^{-1}$ \\
					Far-field stress, $\sigma_\infty$ &  $10$ MPa \\
					Reservoir pressure, $p_\infty$ &  $0$ MPa \\
					Reservoir fluid viscosity, $\eta_r$ &  $10^{-3}$ Pa$\cdot$s\\
					Fracturing fluid viscosity, $\eta_f$ &  $10^{-1}$ Pa$\cdot$s \\
					Injection rate per unit height, $2\,Q$ &  $10^{-3}$ m$^2/$s\\
					\hline
					
				\end{tabular}

			\end{center}
		\end{table}
	In order to verify the algorithm we choose physical parameters typical for hydraulic fracturing problem and check the numerical convergence.
	For the verification we assume the reservoir to be homogeneous. 
	All common parameters for  physical simulations are listed in Table \ref{tab:sim_parameters_verification}. Given Young's modulus $E$ and Poisson's ratio $\nu$, elastic moduli $\lambda$ and $\mu$ are calculated via known formulae
	
\[\lambda=\frac{\nu E}{(1+\nu)(1-2\nu)},\quad \mu = \frac{E}{2(1+\nu)}.\] 

\begin{figure}[h]
		\centering
		\includegraphics[width=0.7\linewidth]{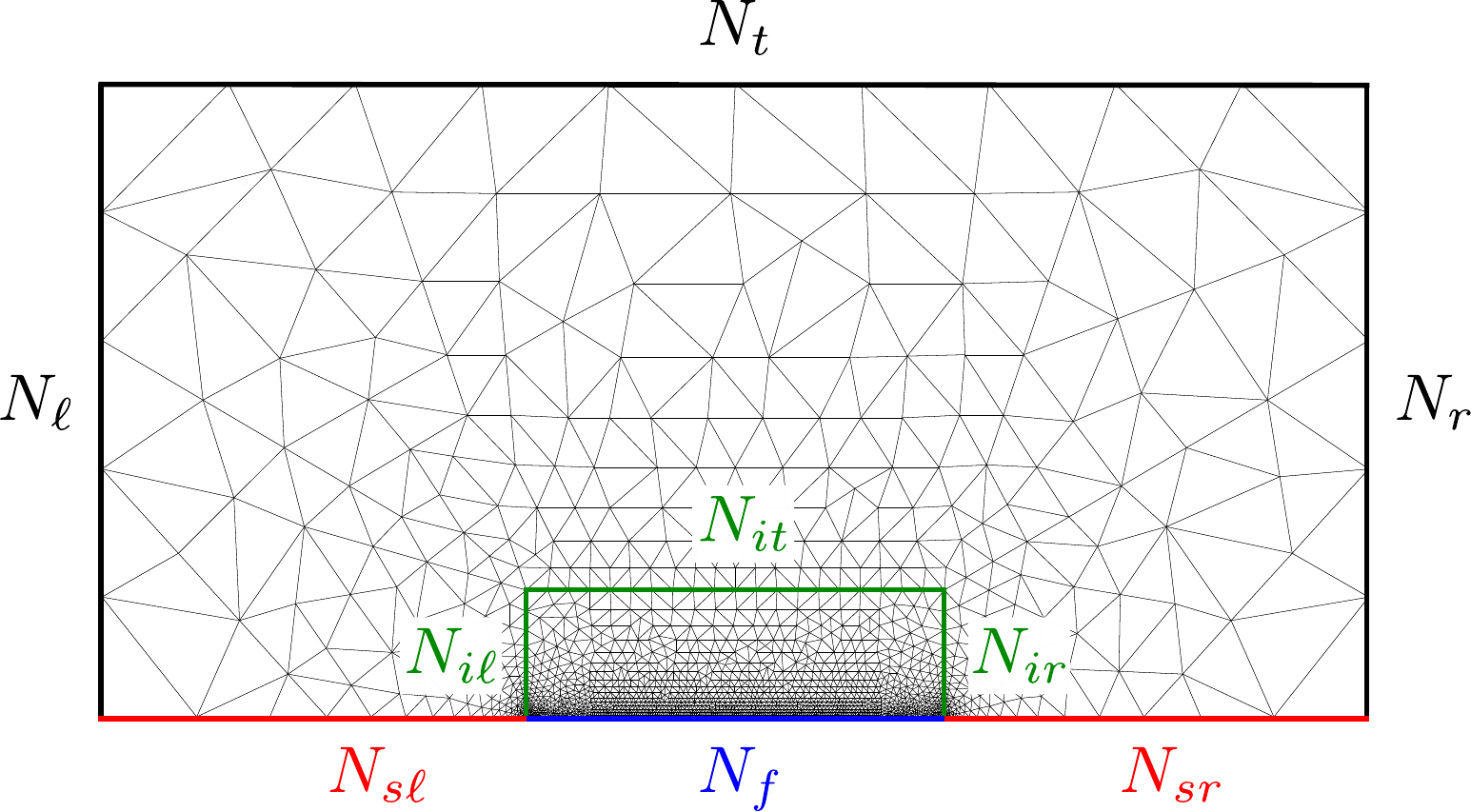}
		\caption{Computational domain triangulation}
		\label{fig:mesh_N100_article}
	\end{figure}

The expression for the storage coefficient $S_\varepsilon$ is given by formula \eqref{eq:storativity_def}.	
	The computational domain was triangulated  using the embedded tool of FreeFEM++  as shown in Fig. \ref{fig:mesh_N100_article}. Here
	\begin{equation} 
		\begin{array}{c}
			N_t = \dfrac{3 N}{400}, \quad N_{\ell} =  N_r = \dfrac{3N}{800}, \\[2ex]
			N_{s\ell} = \dfrac{N}{20} \biggr(\dfrac{R-L_\ell^{max}}{R}\biggl), \quad 	N_{sr} = \dfrac{N}{20} \biggr(\dfrac{R-r^{max}}{R}\biggl), \quad N_{f} = 4N \biggr(\dfrac{L_\ell^{max}+L_r^{max}}{R}\biggl),\\[2ex]
			N_{it} = \dfrac{N}{20} \biggr(\dfrac{L_\ell^{max}+L_r^{max}}{R}\biggl), \quad N_{i\ell} =   N_{ir} = \dfrac{N}{240} 
		\end{array}
	\end{equation}
	denote the number of mesh vertices over the corresponding boundary parts. In order to improve the stability of the algorithm, we redefine the boundary as $\Gamma_s = \{ -R<x<-L_\ell^{max},\; y=0\}\bigcup \{ L_r^{max}<x<R,\; y=0\}$, where $-L_\ell^{max}$ and $L_r^{max}$ are the limiting positions of the corresponding fracture's tips.
	Several simulations were conducted on the sequence of refining meshes with  $N=800\cdot 2^{k-1}$, where  $k=1,\ldots,5$ is the simulation number. 
	
\begin{figure}[!htb]
	\centering
	\includegraphics[width=0.6\linewidth]{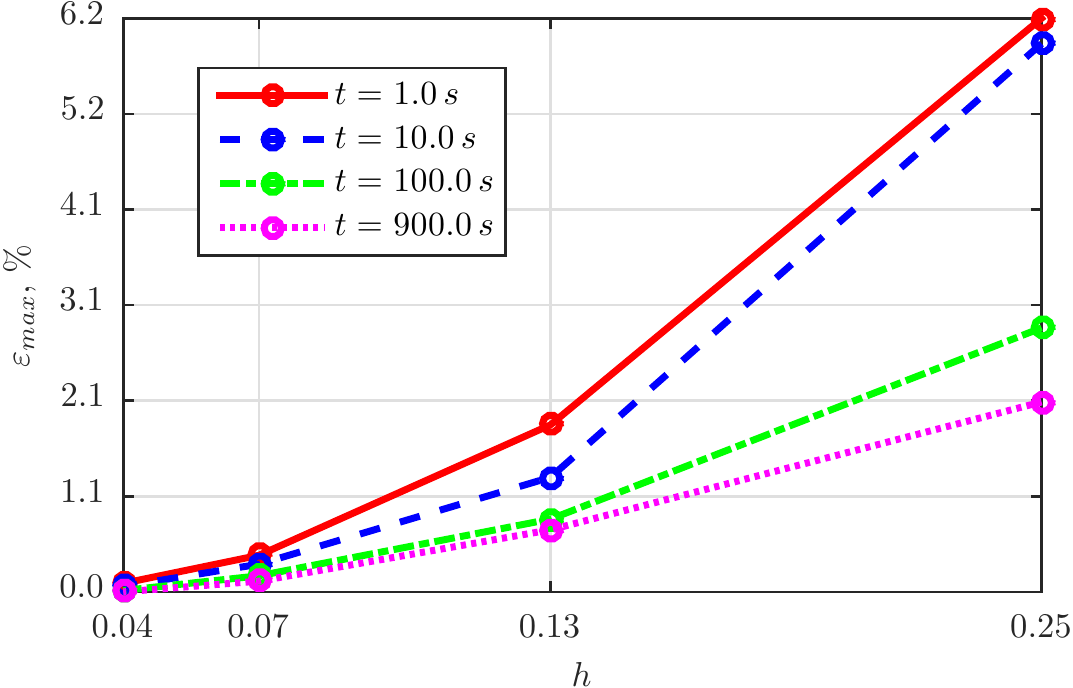}
	\caption{Relative difference in $L_2$-norm  between solutions on two successive meshes for different time moments}
	\label{fig:errorRelativeMaxSeveralTimestepsNumH}
	\end{figure}

	For the convergence test we compute the maximal relative difference in $L_2$-norm between solutions on two successive meshes:	
	\begin{equation}
		\varepsilon_{max}(h)=\max{ \Bigl(\dfrac{||p_{h}-p_{h/2}||_{L_2}}{||p_{h}||_{L_2}},\frac{||u_{h}-u_{h/2}||_{L_2}}{||u_{h}||_{L_2}},\frac{||v_{h}-v_{h/2}||_{L_2}}{||v_{h}||_{L_2}}\Bigr)}\times 100\%.
	\end{equation}
	The result of computations is demonstrated in Fig.  \ref{fig:errorRelativeMaxSeveralTimestepsNumH}, where $h=\sqrt{S_\mathrm{max}}$ is a mesh refining parameter and  $S_\mathrm{max}$ is the maximal dimensionless area of all triangles in the corresponding mesh. One can see that at $h\approx 0.13\,(N=1600)$ the relative difference between solutions is less than 2~\%. We assume this mesh as suitable for engineering purposes and use it for  further simulations.
	
	\subsection{Comparison with existing models}

	According to \cite{Adachi_Plane_Strain_Permeable_Rock,Bunger_Detournay_Garagash} in case of KGD-model the fracture propagation is governed by two competing energy dissipation mechanisms (viscous dissipation and creation of new fracture surface) and two storage mechanisms (in the fracture or in the reservoir). Therefore, there exists four asymptotic regimes: storage-toughness dominated, leakoff-toughness dominated, storage-viscosity dominated and leakoff-viscosity dominated regimes. Using an analogous model of a hydraulic  fracture in a poroelastic media all these regimes were reproduced and showed a good agreement with analytical solutions of KGD-model in \cite{Carrier_Granet}. As a part of model verification we compare our  results with \cite{Carrier_Granet}.

	The common input parameters used in all simulations in this section are the same as  in Table~\ref{tab:sim_parameters_verification} except for the domain size $R = 45$~m, maximal right and left tip positions $L_r^{max} = L_l^{max}=15$ m and some other parameters specified below in each particular case.

	For the storage-toughness dominated regime we use equal viscosities  of reservoir  and fracturing fluids  $\eta_r=\eta_f = 10^{-4}$ Pa$\cdot$s to minimize dissipation due to viscosity. Far-field stress is equal to $\sigma_\infty = 3.7$ MPa. Two simulations were carried out for permeabilities $k_r = 10^{-15}$ m$^2$ and $k_r = 10^{-16}$ m$^2$ during $14$ s and $20$ s, respectively, to ensure that the fracture propagates in the storage regime. According to \cite{Carrier_Granet}, for this set of parameters the results are in a good agreement with early-time near-$K$ solution \cite{Bunger_Detournay_Garagash} for KGD-model.
	
\begin{figure}[h]
		\centering
		\includegraphics[width=0.6\linewidth]{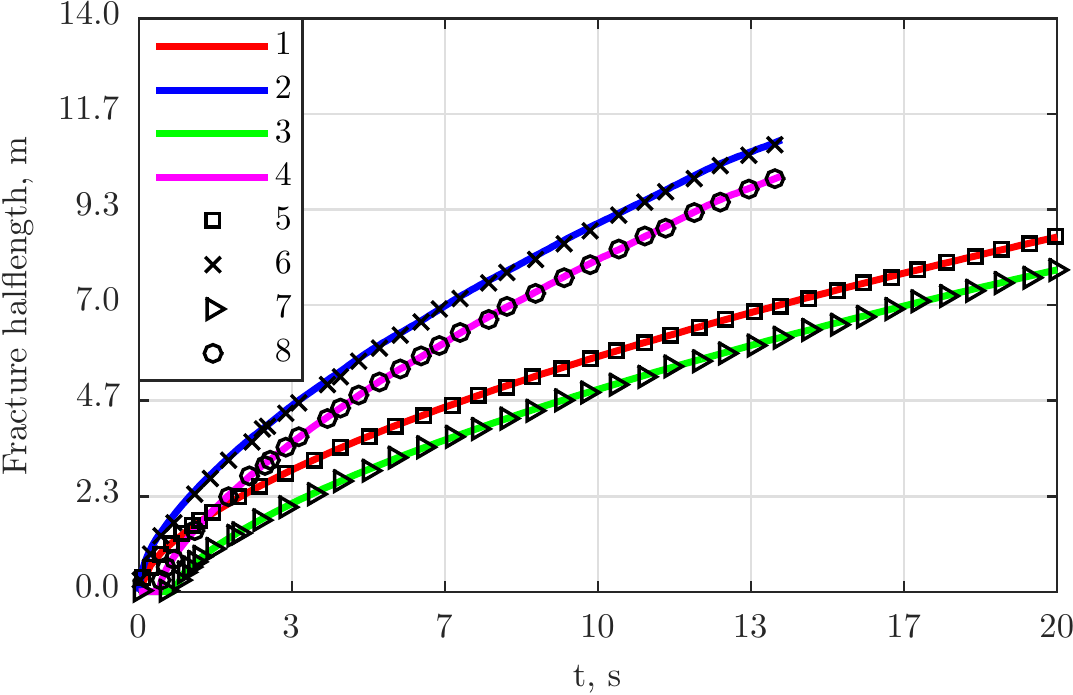}
		\caption{Fracture half-length: for $k_r= 10^{-15}$ m$^2$ (1) in present work, (5) in \cite{Carrier_Granet}; for $k_r= 10^{-16}$ m$^2$  (2)  in present work, (6) in \cite{Carrier_Granet}.  Cohesive zone right end: for $k_r= 10^{-15}$ m$^2$ (3) in present work, (7) in \cite{Carrier_Granet}; for $k_r= 10^{-16}$ m$^2$  (4)  in present work, (8) in \cite{Carrier_Granet}. The regime is  storage-toughness dominated }
		\label{fig:fracHalfLengthStorageToughness7a}
	\end{figure}

\begin{figure}[h]
		\begin{minipage}[t]{0.5\linewidth} 
		
			\includegraphics[width=1\linewidth]{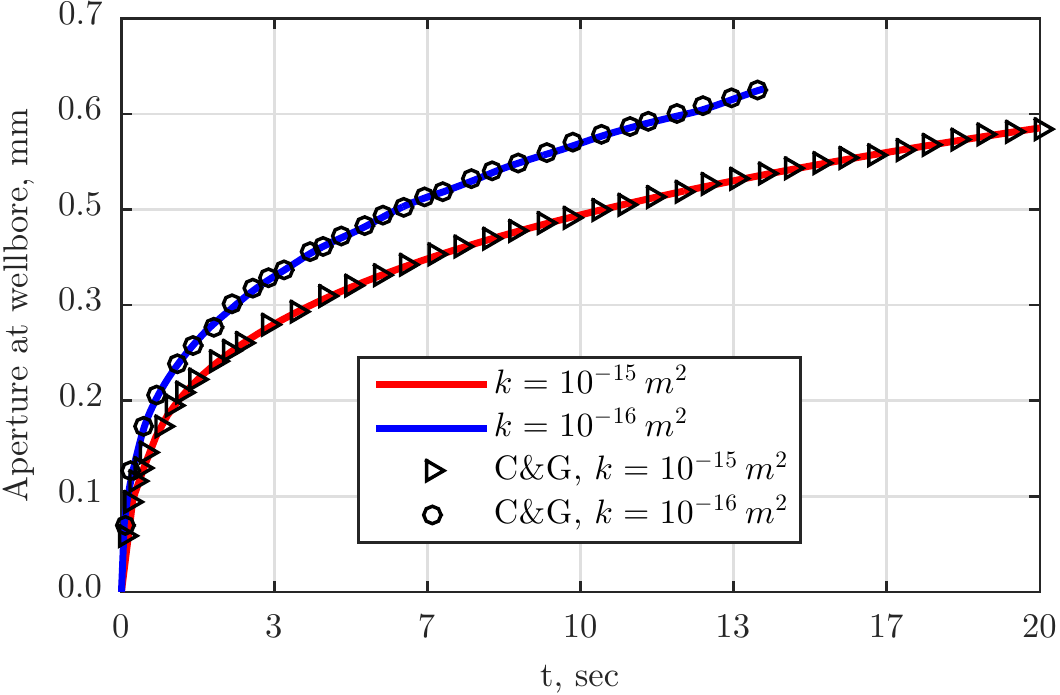}
			\caption{Fracture aperture at the borehole for  $k_r= 10^{-15}$ m$^2$ and  $k_r= 10^{-16}$ m$^2$ in present work and in \cite{Carrier_Granet} (C\&G). The regime is  storage-toughness dominated }
			\label{fig:apertOrigStorageToughness8a}
		
		\end{minipage}
		\hspace{0.5cm}
		\begin{minipage}[t]{0.5\linewidth}
			\includegraphics[width=1\linewidth]{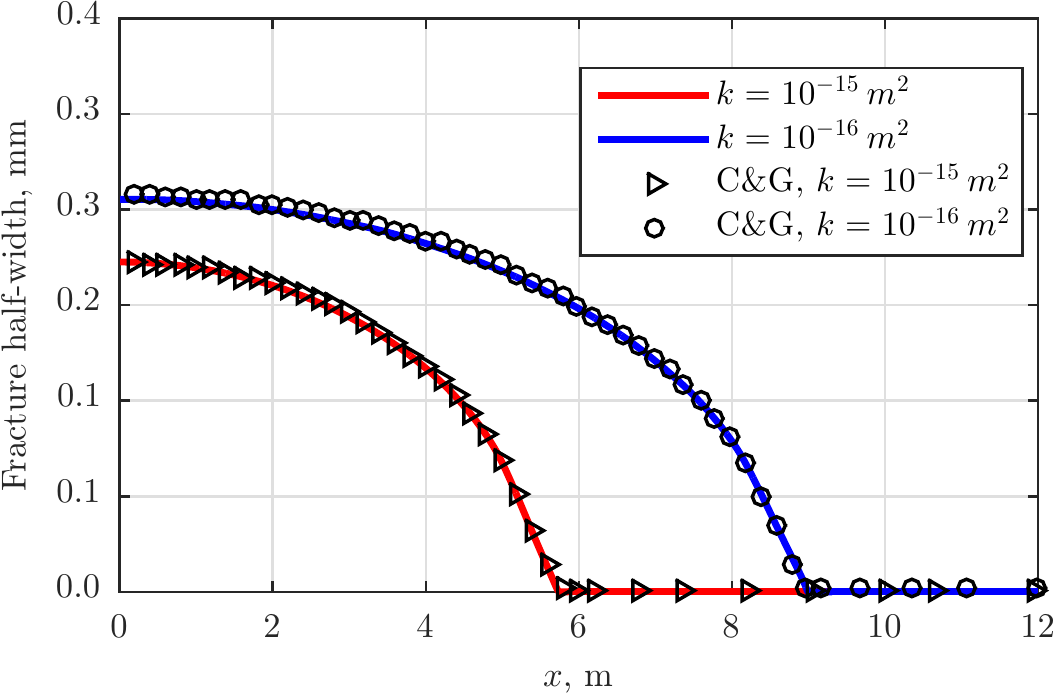}
			\caption{Fracture half-width at $t=10$ s  for  $k_r= 10^{-15}$ m$^2$ and  $k_r= 10^{-16}$ m$^2$ in present work and in \cite{Carrier_Granet} (C\&G). The regime is  storage-toughness dominated }
			\label{fig:fracHalfwidthStorageToughness9a}
		\end{minipage}
	
	\end{figure}

	Figure \ref{fig:fracHalfLengthStorageToughness7a} shows a  coincidence of the fracture half-length in present work (solid lines 1, 2) and in \cite{Carrier_Granet} (marker lines 5, 6). A good match of the end of the cohesive zone (solid lines 3, 5), where the cohesive forces ends, indicates a compliance in the implementation of the cohesive zone model. Figures \ref{fig:apertOrigStorageToughness8a} and \ref{fig:fracHalfwidthStorageToughness9a} show time and space consistency of the fracture geometry with the one obtained in \cite{Carrier_Granet}.

\begin{figure}[h]
		\centering
		\includegraphics[width=0.6\linewidth]{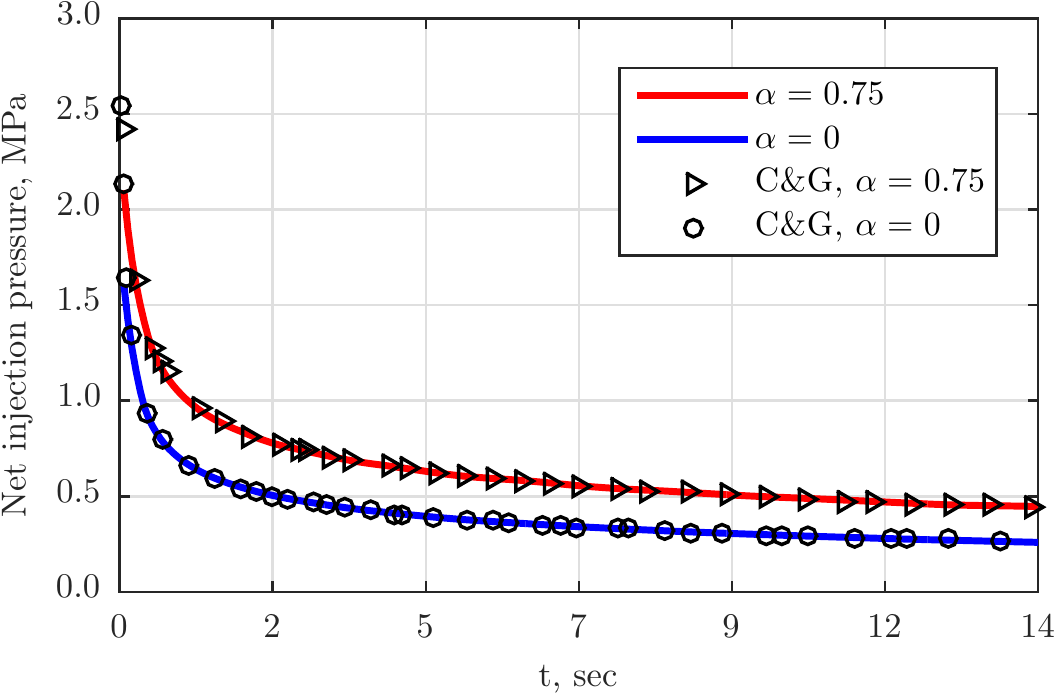}
		\caption{Net injection pressure in storage-toughness dominated regime for coupled ($\alpha =0.75$) and uncoupled ($\alpha =0.0$) cases in the present work and in \cite{Carrier_Granet} (C\&G) for reservoir permeability $k_r= 10^{-16}$ m$^2$} 
		\label{fig:injectionPressureStorageToughness10a}
	\end{figure}

	Keeping the storage coefficient $S_\varepsilon$ fixed  we can impose $\alpha=0$ thus removing the coupling between elastic  \eqref{eq:variat_formul_equil} and filtration \eqref{eq:variat_formul_filtr} equations. Comparison of the net pressure  $p_{net} = p - \sigma_\infty$ is presented in Figure \ref{fig:injectionPressureStorageToughness10a}.

\begin{figure}[h]
		\begin{minipage}[t]{0.5\linewidth} 
			\includegraphics[width=1\linewidth]{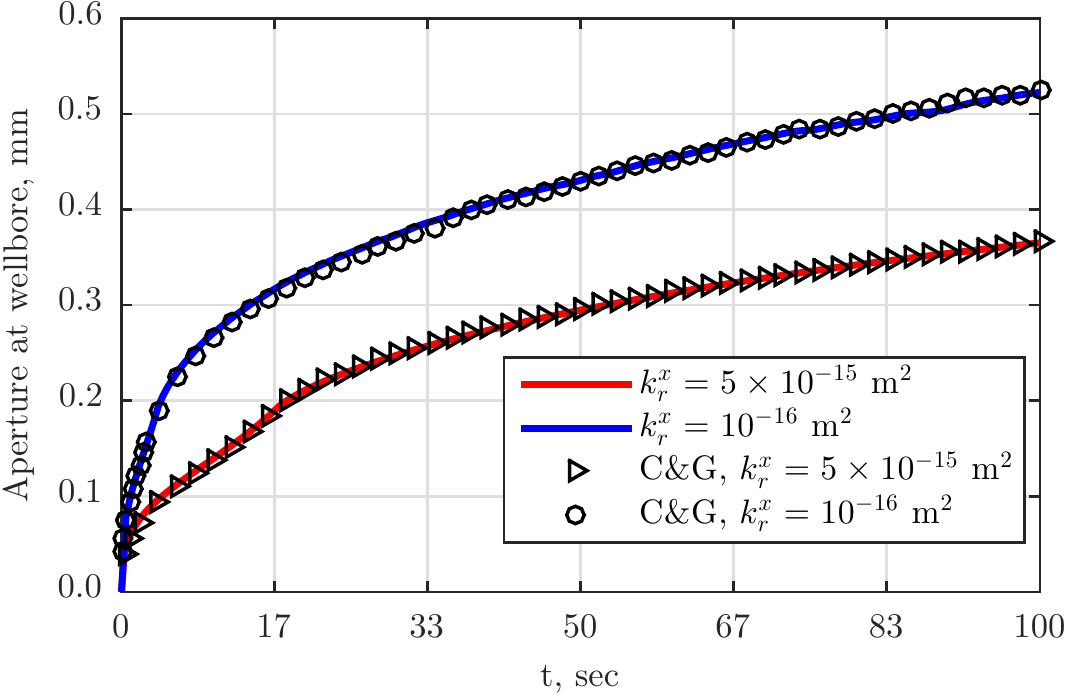}
			\centering{\bf(a)}	
		\end{minipage}
		\hspace{0.5cm}
		\begin{minipage}[t]{0.5\linewidth}
			
			\includegraphics[width=1\linewidth]{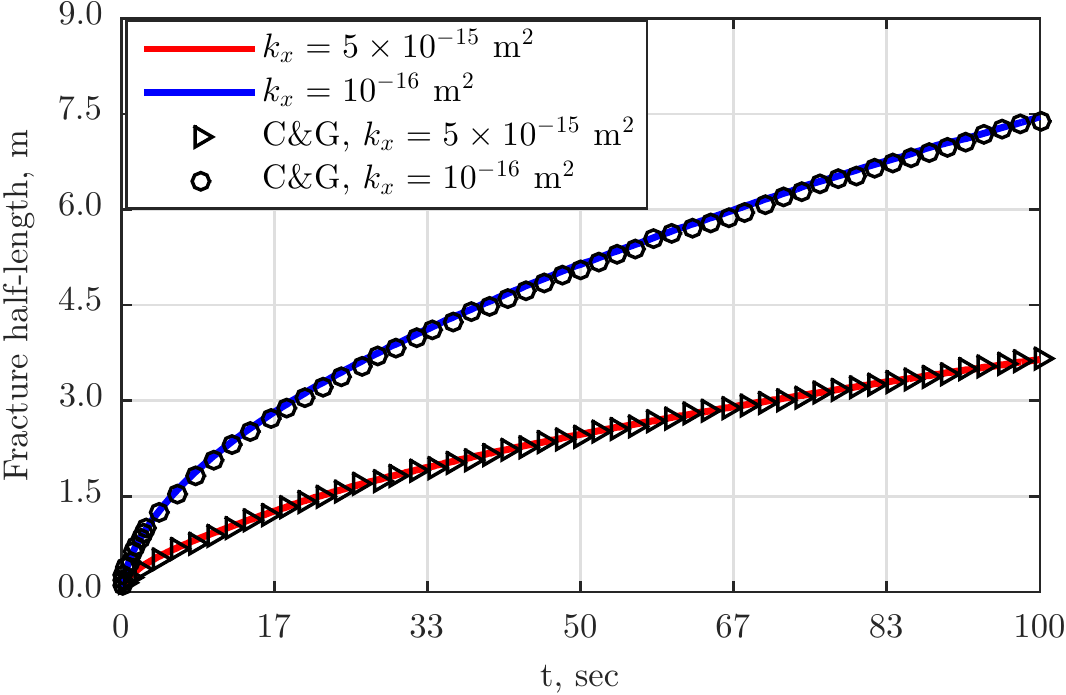}
				
			\centering{\bf(b)}
		\end{minipage}
		\caption{Fracture aperture (a) and half-length (b) for two-dimensional ($k_r^x = 5\times10^{-15}$~m$^2$) and unidimensonal leakoff ($k_r^x =  10^{-16}$ m$^2$) in the present work and in \cite{Carrier_Granet} (C\&G)} for the leakoff-toughness dominated regime
		\label{fig:apertAndHalfLengthLeakoffToughness}
	\end{figure}

 In Figure \ref{fig:apertAndHalfLengthLeakoffToughness} we compare the fracture width at the borehole (a) and the fracture half-length~(b) with the same data in \cite{Carrier_Granet} for the leakoff-toughness dominated regime. The fluid viscosities remain the same as in the storage-toughness regime. The far-field stress $\sigma_\infty = 5$ MPa and the permeability in $y$-direction $k_r^y = 5\times10^{-15}$ m$^2$ are increased providing the leakoff to be large enough to reach the leakoff dominated regime. The permeability in $x$-direction is equal to $k_r^x = k_r^y$ in case of 2D fluid diffusion in the reservoir and equal to $k_r^x = 0.02\,k_r^y$  in case of unidimensional fluid diffusion. In the unidimensional case the numerical solution is agreed with near-$\tilde{K}$ solution \cite{Bunger_Detournay_Garagash}.
	
\begin{figure}[h]
		\begin{minipage}[t]{0.5\linewidth} 
			
			\includegraphics[width=1\linewidth]{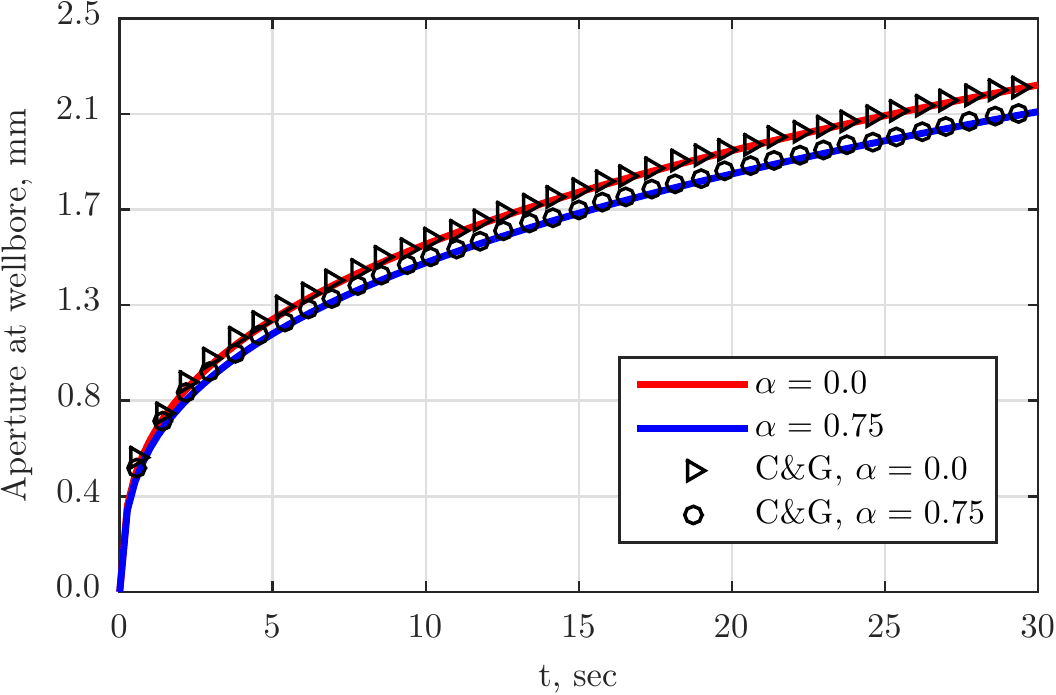}
			
			\centering{\bf(a)}
		\end{minipage}
		\hspace{0.5cm}
		\begin{minipage}[t]{0.5\linewidth}
			
			\includegraphics[width=1\linewidth]{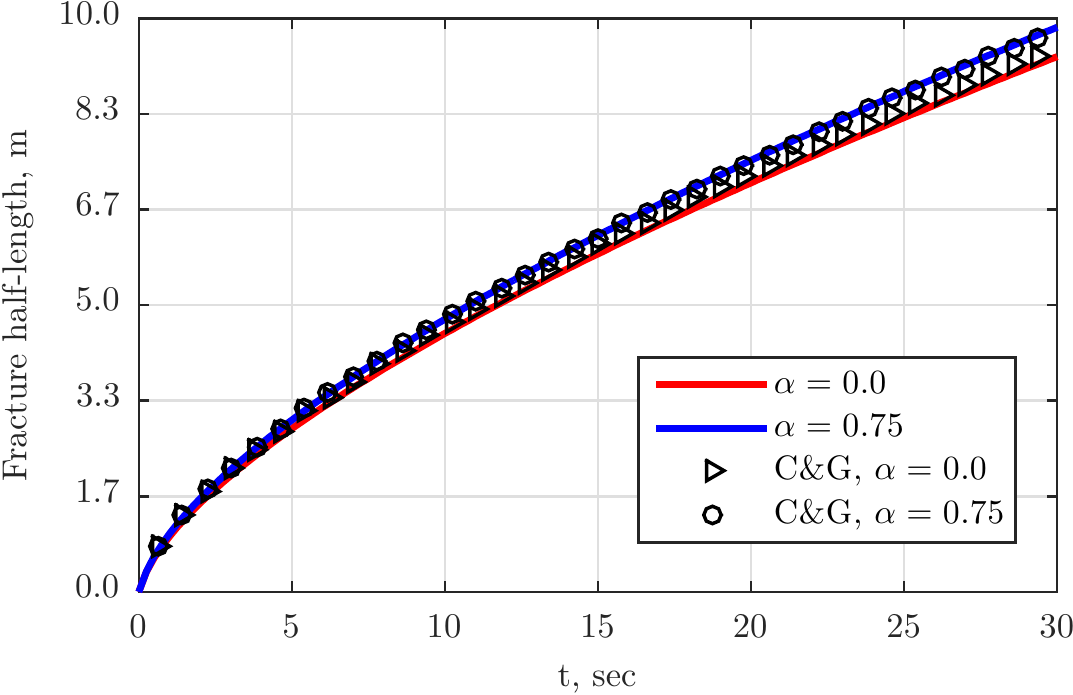}

			\centering{\bf(b)}
		\end{minipage}
		\caption{Fracture aperture (a) and half-length (b) for coupled ($\alpha =0.75$) and uncoupled ($\alpha =0$) cases in the present work and in \cite{Carrier_Granet} (C\&G) in the storage-viscosity dominated regime }
		\label{fig:apertAndHalfLengthStorageToughnessZeroLeakoff}
	\end{figure}

\begin{figure}[h]
		\centering
		\includegraphics[width=0.6\linewidth]{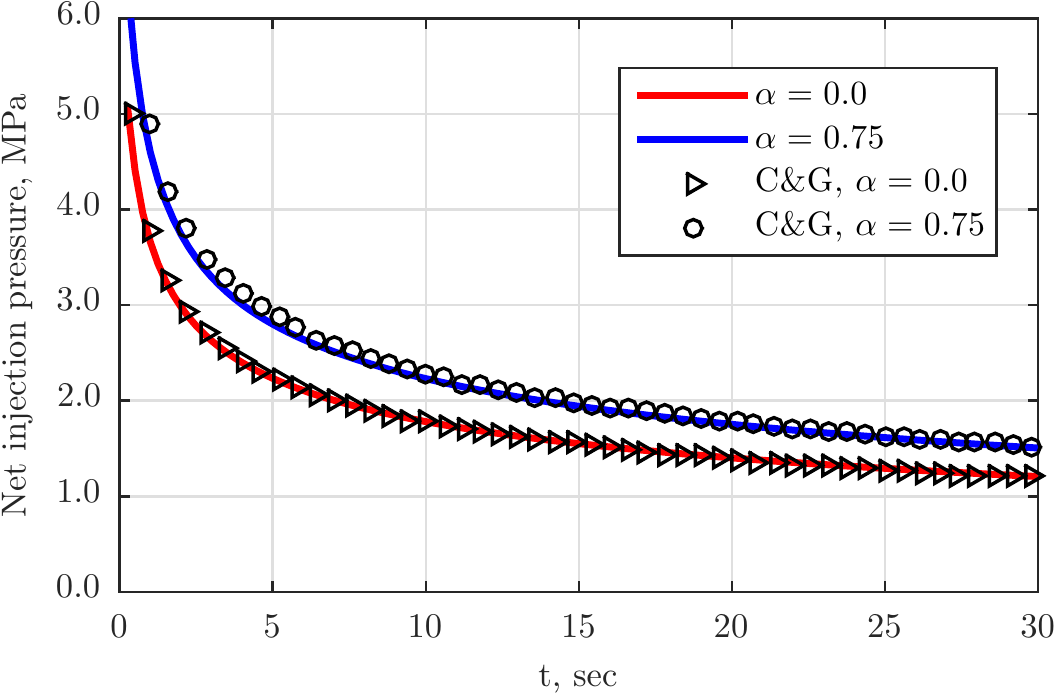}
		\caption{Net injection pressure for coupled ($\alpha =0.75$) and uncoupled ($\alpha =0$) cases in the present work and in \cite{Carrier_Granet} (C\&G) in the storage-viscosity dominated regime}
		\label{fig:pOrigStorageToughnessZeroLeakoff19}
	\end{figure}

To reproduce the storage-viscosity dominated regime, the following fluid viscosities are chosen: $\eta_r=\eta_f = 10^{-1}$ Pa$\cdot$s. The reservoir permeability is taken as  $k_r=10^{-15}$ m$^2$ and the far-field stress is $\sigma_\infty = 3.7$ MPa. Figures \ref{fig:apertAndHalfLengthStorageToughnessZeroLeakoff}, \ref{fig:pOrigStorageToughnessZeroLeakoff19} show coincidence  of the fracture half-length, fracture width and pressure  at the borehole with the data obtained in \cite{Carrier_Granet} for coupled ($\alpha=0.75$) and uncoupled ($\alpha=0$) cases. The corresponding self-similar solution agreed with the uncoupled case is $M$-solution \cite{Adachi_Plane_Strain_Permeable_Rock}. It was pointed out in \cite{Carrier_Granet} that in this case changing Biot coefficient $\alpha$ only slightly affects the fracture geometry, but induces the increase of the fracture pressure due to the so called backstress \cite{Vandamme_Roegiers}, \cite{Kovalyshen_PhD} arising near the fracture walls. The backstress effect will be discussed later in this paper.   
	
\begin{figure}[h]
		\begin{minipage}[t]{0.5\linewidth} 
			
			\includegraphics[width=1\linewidth]{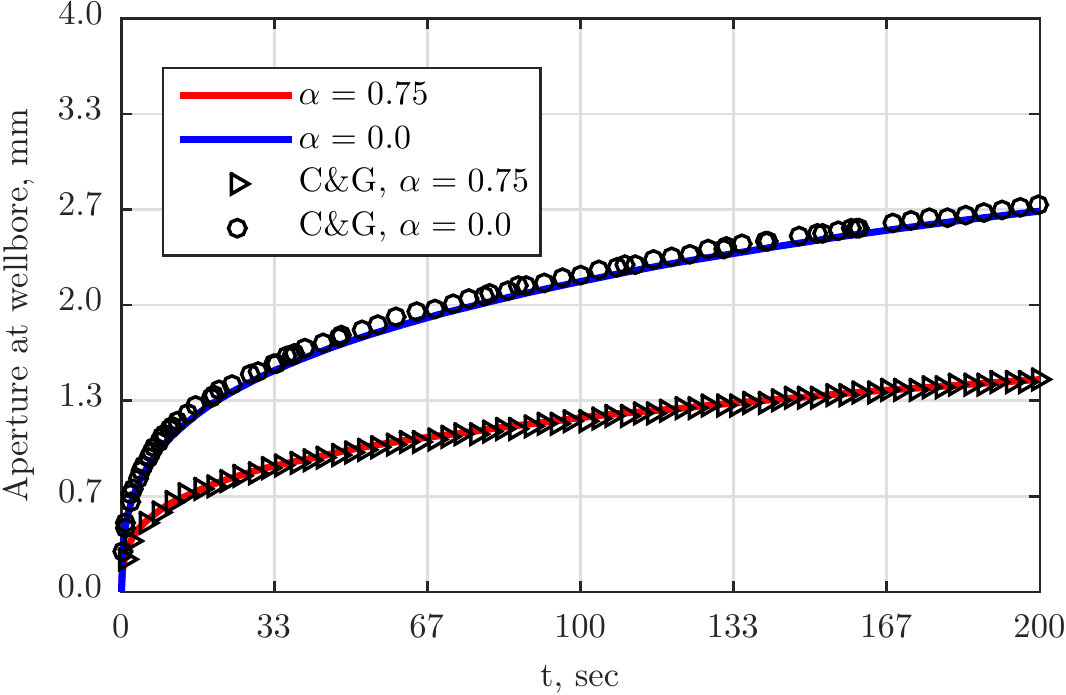}
			
			\centering{\bf(a)}
		\end{minipage}
		\hspace{0.5cm}
		\begin{minipage}[t]{0.5\linewidth}

			\includegraphics[width=1\linewidth]{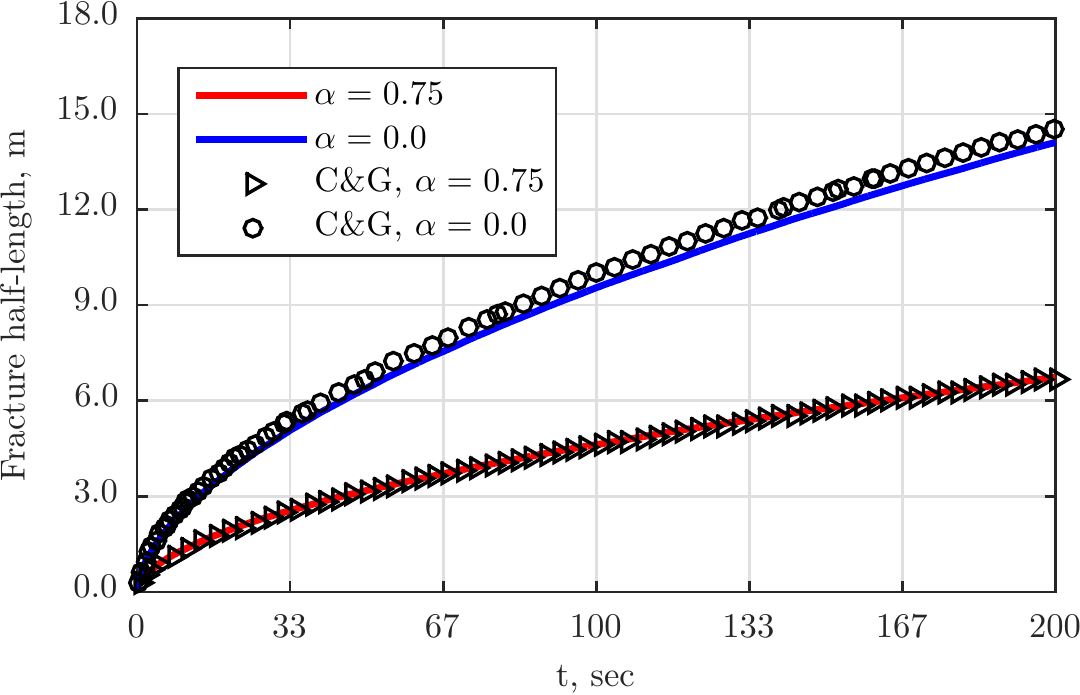}
			
			\centering{\bf(b)}
		\end{minipage}
		\caption{Fracture aperture (a) and half-length (b) for coupled ($\alpha =0.75$) and uncoupled ($\alpha =0.0$) cases for the unidimensional leakoff in the present work and in \cite{Carrier_Granet} (C\&G) in the leakoff-viscosity dominated regime }
		\label{fig:apertAndHalfLengthLeakoffViscosity}
	\end{figure}

\begin{figure}[h]
		\centering
		\includegraphics[width=0.6\linewidth]{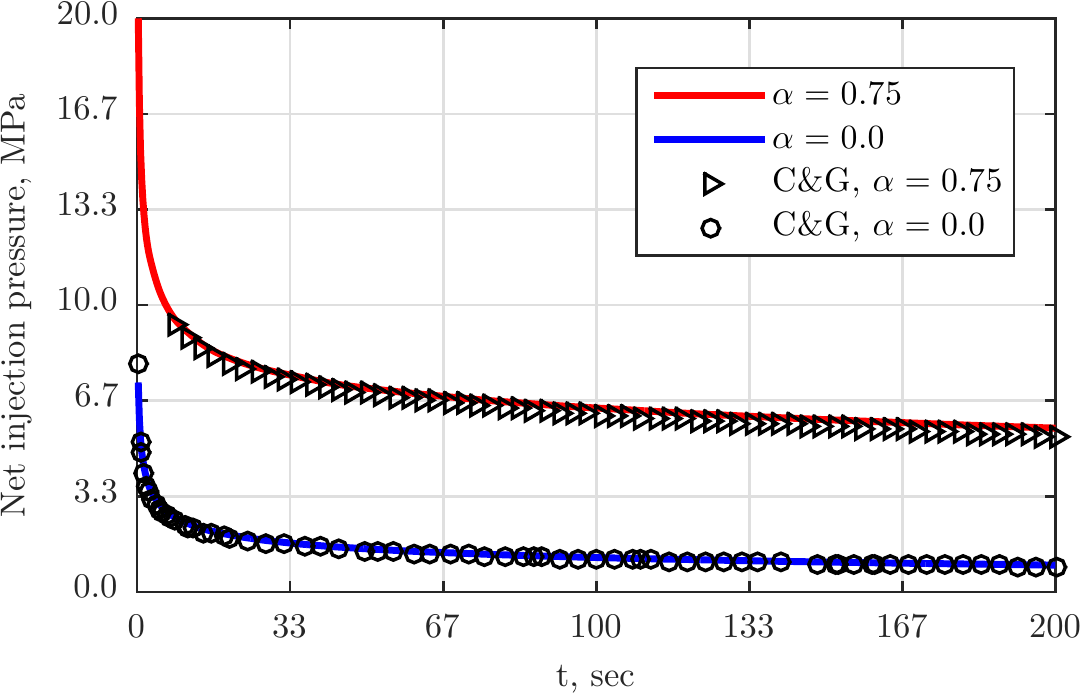}
		\caption{Net injection pressure for the coupled ($\alpha =0.75$) and uncoupled ($\alpha =0.0$) cases for the unidimensional leakoff in the present work and in \cite{Carrier_Granet} (C\&G) in the leakoff-viscosity dominated regime}
		\label{fig:injectionPressureLeakoffViscosity23}
	\end{figure}

	For the leakoff-viscosity dominated regime we perform two simulations with unidimensional fluid diffusion in the reservoir for the coupled and uncoupled cases. The permeability is taken as $k_r^y = 5\times10^{-12}$ m$^2$ in $y$-direction and  is  $k_r^x = 0.00002\,k_r^y$ in $x$-direction. The far-field stress $\sigma_\infty$ is equal to $7.2$ MPa and the fluid viscosities are chosen as $\eta_r=\eta_f = 10^{-1}$ Pa$\cdot$s. Figures \ref{fig:apertAndHalfLengthLeakoffViscosity}, \ref{fig:injectionPressureLeakoffViscosity23} show the fracture width at the borehole, fracture half-length and net injection pressure $p_{net}$, respectively, in comparison to the data from \cite{Carrier_Granet}. All curves in the uncoupled case match the $\tilde{M}$ analytical solution \cite{Adachi_Plane_Strain_Permeable_Rock} for KGD-model. In the coupled case the pressure in the fracture is significantly larger due to the poroelastic back-stress. In turn, it results in larger value of leakoff and shorter fracture.

	\section{Influence of the pore pressure to the fracture dynamics}

In this section we use the developed model in order to study the influence of the pore pressure $p$ to the distribution of stresses in the vicinity of the fracture and to the consequent dynamics of the fracture propagation. The common physical parameters used in numerical experiments are listed in Table \ref{tab:sim_parameters_verification}. 

\subsection{The Biot's number}
\begin{figure}[h]
		\begin{minipage}[t]{0.5\linewidth} 
			
			\includegraphics[width=1\linewidth]{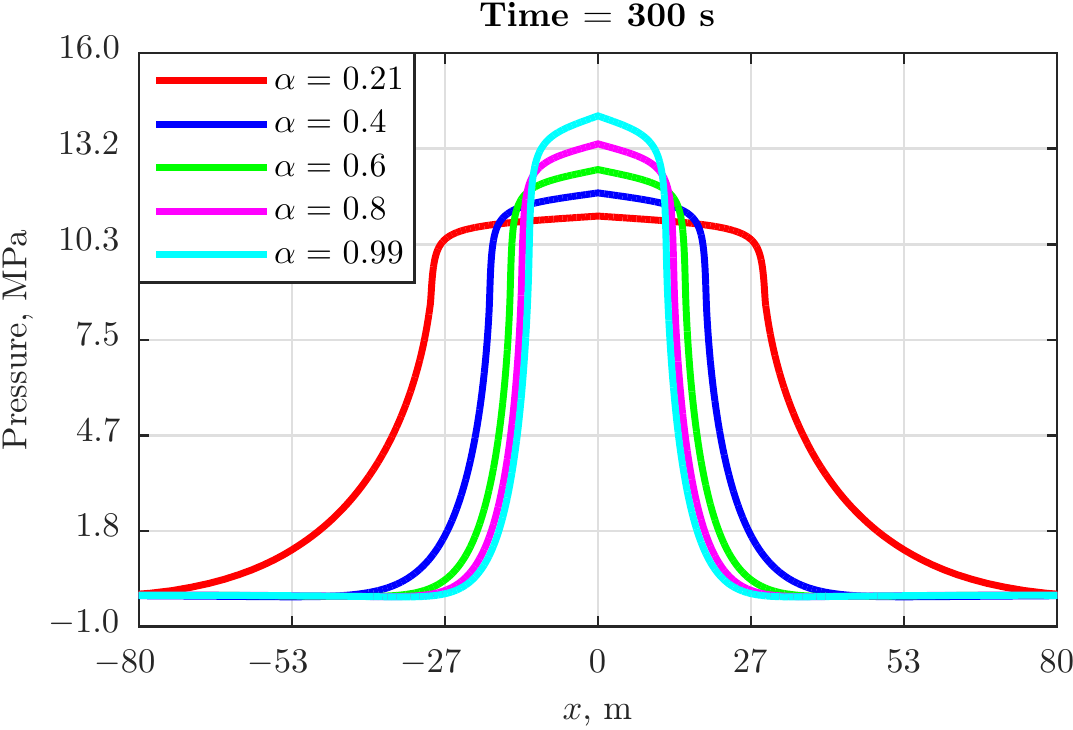}
			\centering {\bf(a)}

		\end{minipage}
		\hspace{0.5cm}
		\begin{minipage}[t]{0.5\linewidth}
			
			\includegraphics[width=1\linewidth]{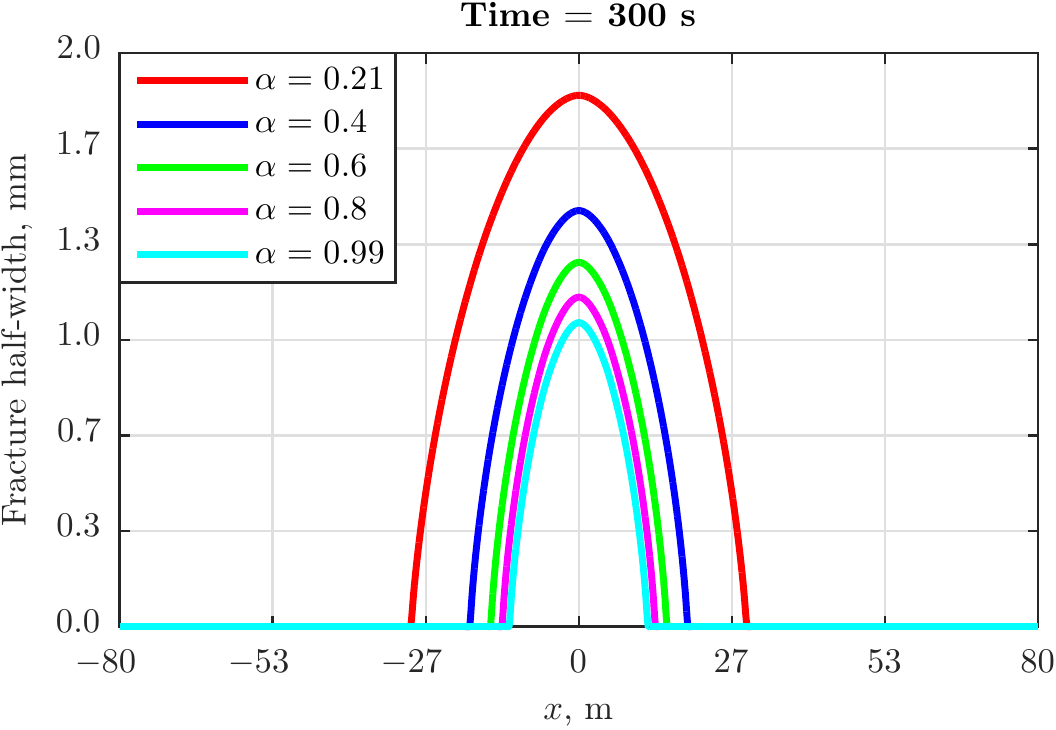}
			\centering{\bf(b)}
			
		\end{minipage}
		\caption{Pressure (a) and fracture half-width (b) along the fracture for different values of Biot's coefficient $\alpha$ at $t=300$ s}
		\label{fig:widthAndPressureProfileOnAlphaTimestep200}
	\end{figure}

The Biot number $\alpha$ is a parameter that determines the contribution of the pore pressure to the total stress (see eq. \eqref{eq:model}). Zero Biot's number $\alpha = 0$ implies that the pore pressure and the elastic stress are decoupled so that the filtration process and the rock deformation are not interrelated. The highest Biot's number $\alpha = 1$ implies the maximal coupling of the pore pressure and the stress. The general influence of the pore pressure $p$ and the dependence of fracture parameters (length and width) on the Biot's coefficient are demonstrated in Figure~\ref{fig:widthAndPressureProfileOnAlphaTimestep200}. It can be seen that for greater $\alpha$ the pressure inside the fracture is higher (see Figure~\ref{fig:widthAndPressureProfileOnAlphaTimestep200}~(a)) whereas the fracture is shorter (see Figure~\ref{fig:widthAndPressureProfileOnAlphaTimestep200}~(b)). 

This effect is the interplay of two factors: change of the stress distribution near the fracture due to the additional hydrostatic compression of the rock by the pore fluid, and influence of the rock deformation to the filtration of fluid. In the forthcoming analysis we separate these two factors and demonstrate the contribution of each to the fracture dynamics.

In order to conduct the thorough study of the influence of the pore pressure and the fluid filtration to the fracture propagation, we distinguish the Biot's number $\alpha$ in equilibrium equation~\eqref{eq:tild_formul_equilib} and in filtration equation~\eqref{eq:tild_formul_filtr} by denoting it as $\alpha_e$ and $\alpha_f$ respectively. So that the case $\alpha_e=0$, $\alpha_f\ne0$ implies zero impact of the pore pressure to the stress, whereas the case $\alpha_e\ne0$, $\alpha_f=0$ corresponds to the uncoupled rock deformation and fluid filtration processes.

\begin{figure}[h]
	\begin{minipage}[t]{0.5\linewidth} 
		
		\includegraphics[width=1\linewidth]{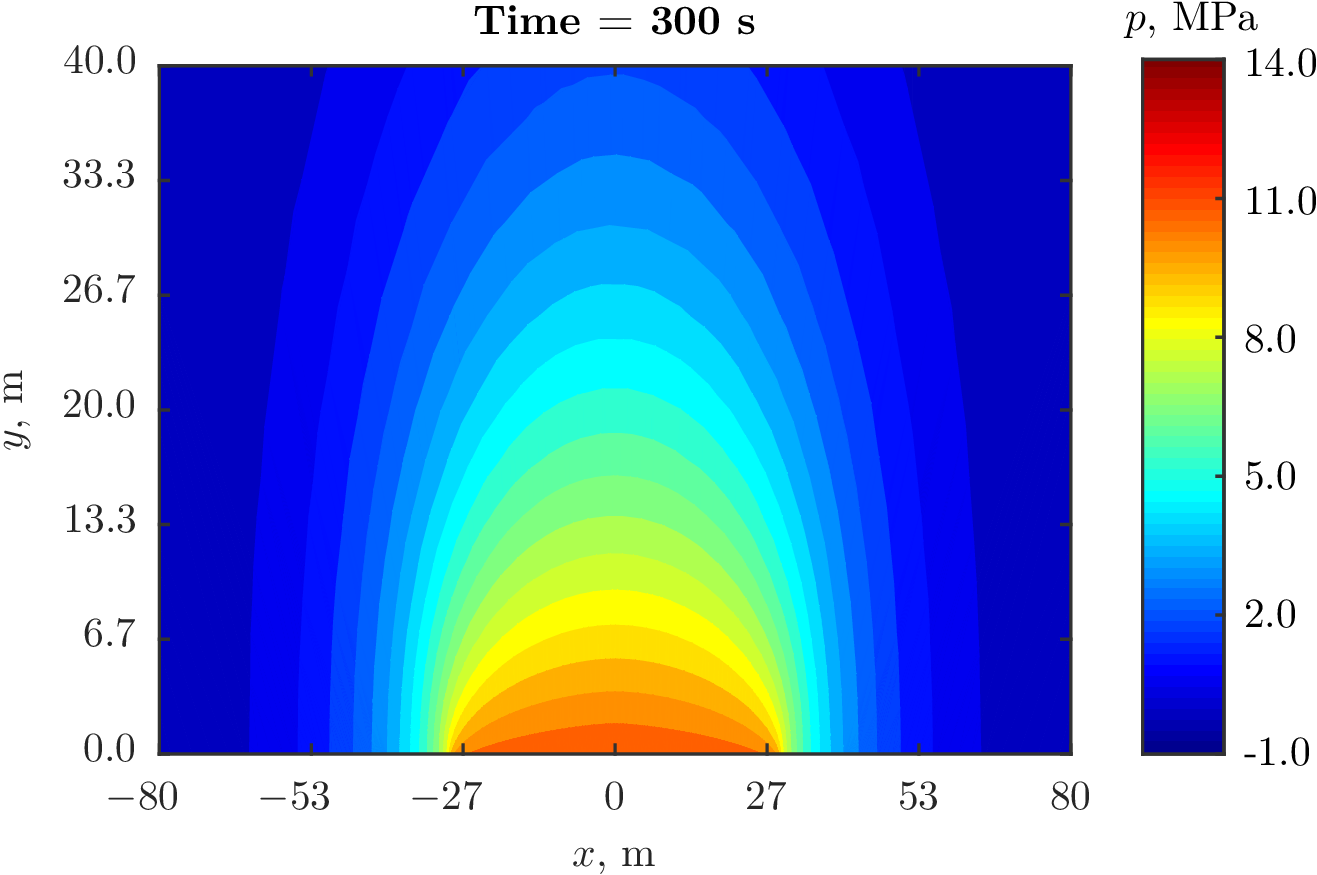}
		\centering {\bf(a)}
		
	\end{minipage}
	\hspace{0.5cm}
	\begin{minipage}[t]{0.5\linewidth}
		
		\includegraphics[width=1\linewidth]{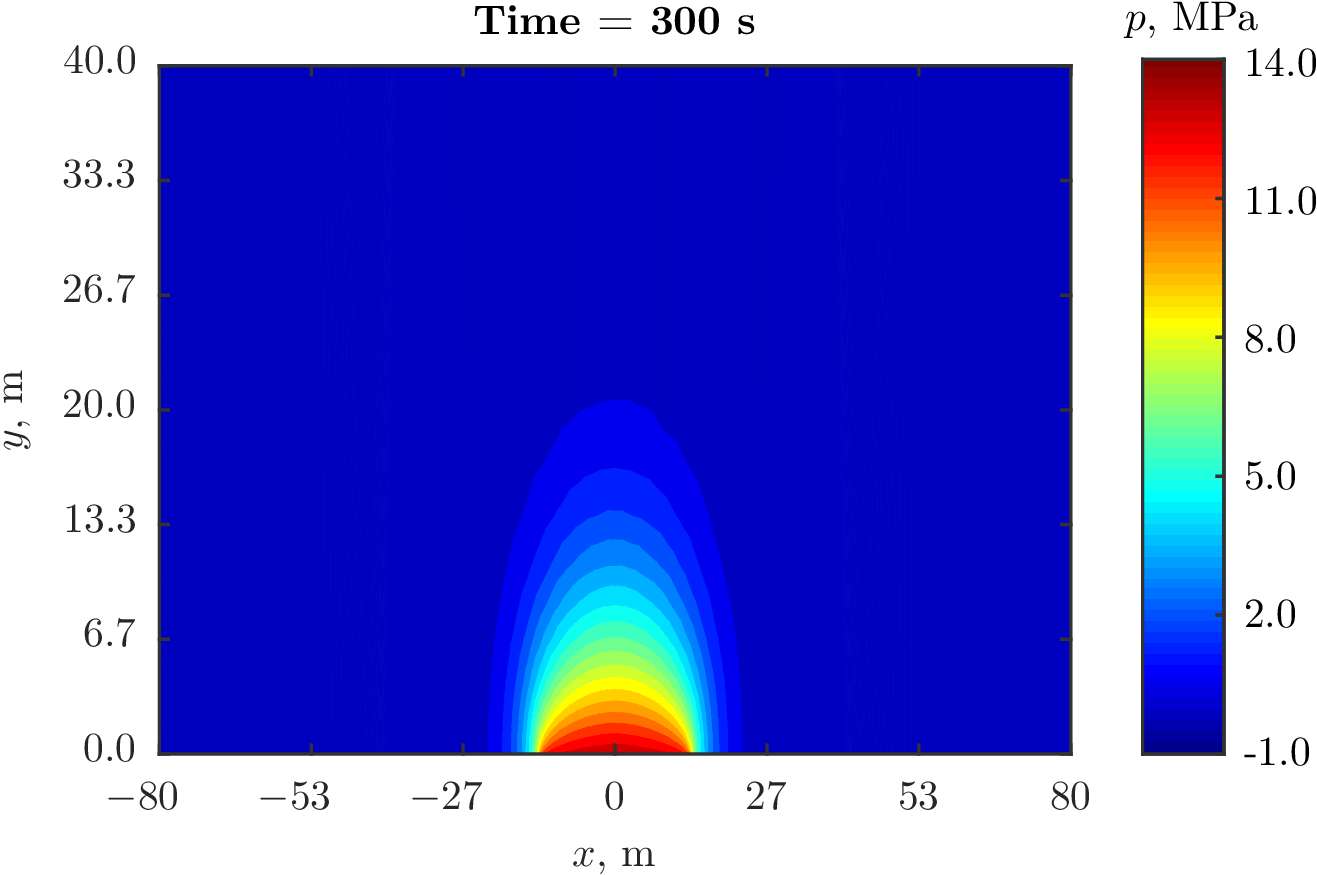}
		\centering{\bf(b)}
		
	\end{minipage}
	
	\caption{Fluid pressure distribution in the reservoir for $\alpha=0.21$ (a) and $\alpha=0.99$ (b)  at $t=300$ s}
	\label{fig:pressure3DAlphaTimestep200}
	
\end{figure}

Fixing the storativity $S_\varepsilon$ and permeability $k_r$, we compare the simulation results in four cases: A) $\alpha_e=0.75$, $\alpha_f=0.75$ (fully coupled); B) $\alpha_e=0.75$, $\alpha_f=0.0$ (partially coupled); C) $\alpha_e=0.0$, $\alpha_f=0.75$ (partially coupled); D) $\alpha_e=0.0$, $\alpha_f=0.0$ (uncoupled).  Figure~\ref{fig:pressureAndWidthProfileAlphaZeroNonZeroHiPermLoSeTimestep200} shows the pressure and the half-width profiles along the fracture path in the case of high reservoir permeability ($k_r = 10^{-14}$ m$^2$) and low storativity ($S_\varepsilon = 1.46\times 10^{-11}$~Pa$^{-1}$) at time $t=300$~s. Note that the shortest fracture is obtained in the fully coupled case. As it will be shown in the forthcoming sections, the reduction of the fracture length in the fully coupled case is the consequence of higher fluid leakoff from the fracture as it can be seen in Figure~\ref{fig:pressure3DAlphaTimestep200} where we compare pressure distribution in the reservoir in cases of low and high Biot's number $\alpha$.

\subsection{Influence of the rock deformation to the fluid filtration}\label{FirstThumbRule}

\begin{figure}[h]
	\begin{minipage}[t]{0.5\linewidth} 
		
		\includegraphics[width=1\linewidth]{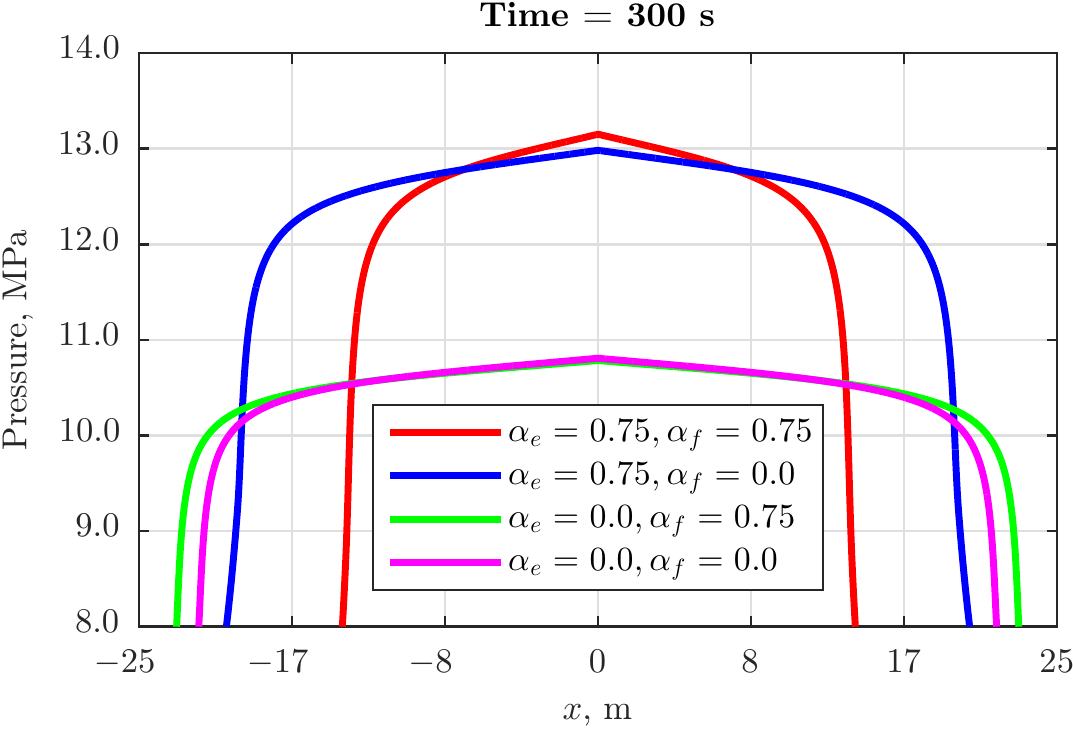}
		\centering {\bf(a)}
		
	\end{minipage}
	\hspace{0.5cm}
	\begin{minipage}[t]{0.5\linewidth}
		
		\includegraphics[width=1\linewidth]{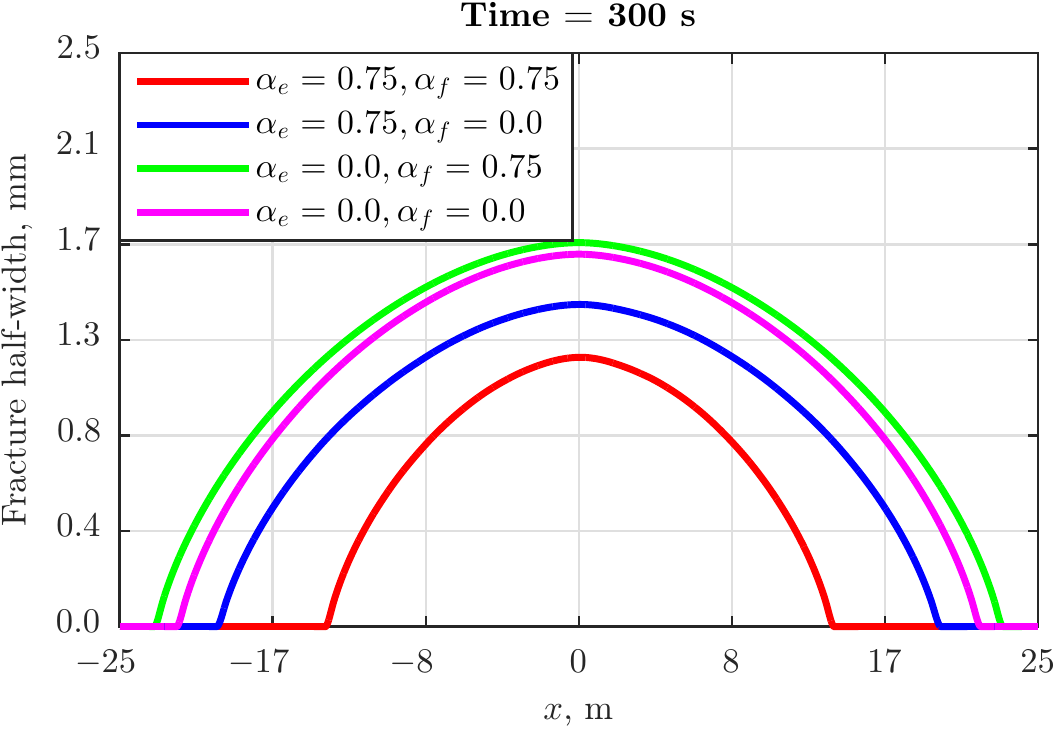}
		\centering{\bf(b)}
		
	\end{minipage}
	
	\caption{Pressure (a) and fracture half-width (b) along the fracture for the coupled, uncoupled and partially coupled cases at $t=300$ s. The physical parameters correspond to relatively high permeability and low storativity:  $k_r = 10^{-14}$ m$^2$, $S_\varepsilon = 1.46\times 10^{-11}$~Pa$^{-1}$.}
	\label{fig:pressureAndWidthProfileAlphaZeroNonZeroHiPermLoSeTimestep200}
	
\end{figure}

In order to distinguish the influence of the rock deformation to the filtration of fluid we compare cases C) $\alpha_e=0$, $\alpha_f = 0.75$ and D) $\alpha_e=\alpha_f=0$ in Figure~\ref{fig:pressureAndWidthProfileAlphaZeroNonZeroHiPermLoSeTimestep200}. It can be seen that the fracture is slightly shorter in  the uncoupled case D. As it follows from the filtration equation \eqref{eq:tild_formul_filtr}, for zero storage coefficient $S_\varepsilon=0$ the flow rate of fluid through every closed surface is proportional to the negative rate of the surface deformation. This implies, that the rock deformation generates the fluid filtration in the direction opposite to the time derivative of the displacement vector. This effect is dumped by the non-zero material storativity $S_\varepsilon\ne0$.

In application to the cases C and D in Figure~\ref{fig:pressureAndWidthProfileAlphaZeroNonZeroHiPermLoSeTimestep200}, this implies that in case C deformation of rock in the direction perpendicular to the fracture's wall generate the backflow of fluid towards the fracture, that reduces the overall fluid leakoff and, consequently, increases the volume of the obtained fracture. The observations above can be summed up in the First thumb rule: ``deformation causes counter filtration''.

\subsection{Influence of the pore pressure to the stress}
To reveal the action of the pore fluid to the stress near the fracture we compare cases B)~$\alpha_e=0.75$, $\alpha_f = 0$ and D) $\alpha_e=\alpha_f=0$  in Figure~\ref{fig:pressureAndWidthProfileAlphaZeroNonZeroHiPermLoSeTimestep200}. In both cases the rock deformation does not change the fluid filtration. However, in the partially uncoupled case B the pore fluid creates an additional hydrostatic compression of rock that compensates a part of the pressure of fracturing fluid to fracture's walls. This effect is referred to as the backstress \cite{Kovalyshen_PhD,Vandamme_Roegiers}. The backstress effectively reduces the pressure applied to fracture's walls by the fracturing fluid. In turn, this causes the increase of fluid pressure inside the fracture in order to maintain the fracture opening required to accommodate the given fluid inflow at the wellbore. This observation is supported by the comparison of graphs of pressure for cases B and D in Figure~\ref{fig:pressureAndWidthProfileAlphaZeroNonZeroHiPermLoSeTimestep200}. The increase of the pressure inside the fracture leads to the higher fluid loss due to the filtration through fracture's walls, which in turn reduces the effectiveness of fluid and decreases the fracture size (compare graphs for fracture half-width in Figure~\ref{fig:pressureAndWidthProfileAlphaZeroNonZeroHiPermLoSeTimestep200}). The Second thumb rule for the action of the pore fluid to the rock deformation is thus ``the pore pressure stiffens the rock''. 

\subsection{Fully coupled action of pore fluid to the hydraulic fracture dynamics}

In the fully coupled case the fracture propagation is governed by the interplay of the two factors: influence of the rock deformation to the fluid filtration and action of the backstress that rises the pressure of fracturing fluid and increases the leakoff. As it can be seen by comparison of graphs A) $\alpha_e=0.75$, $\alpha_f = 0.75$ and D) $\alpha_e=\alpha_f=0$  in Figure~\ref{fig:pressureAndWidthProfileAlphaZeroNonZeroHiPermLoSeTimestep200}, the fracture is about 20\% wider and longer for the uncoupled case D than in the fully coupled case A. 

The reason for the decrease of fracture's volume is twofold. First, in the coupled case the fluid pressure within the fracture is higher due to the backstress (the Second thumb rule), hence, the leakoff rate is higher as well and the fracture volume is lower. However, this factor alone can not explain the significant difference in the fracture geometry as it follows from the comparison of cases B and D in the previous section.  

\begin{figure}[h]
	\begin{minipage}[t]{0.5\linewidth} 
		
		\includegraphics[width=1\linewidth]{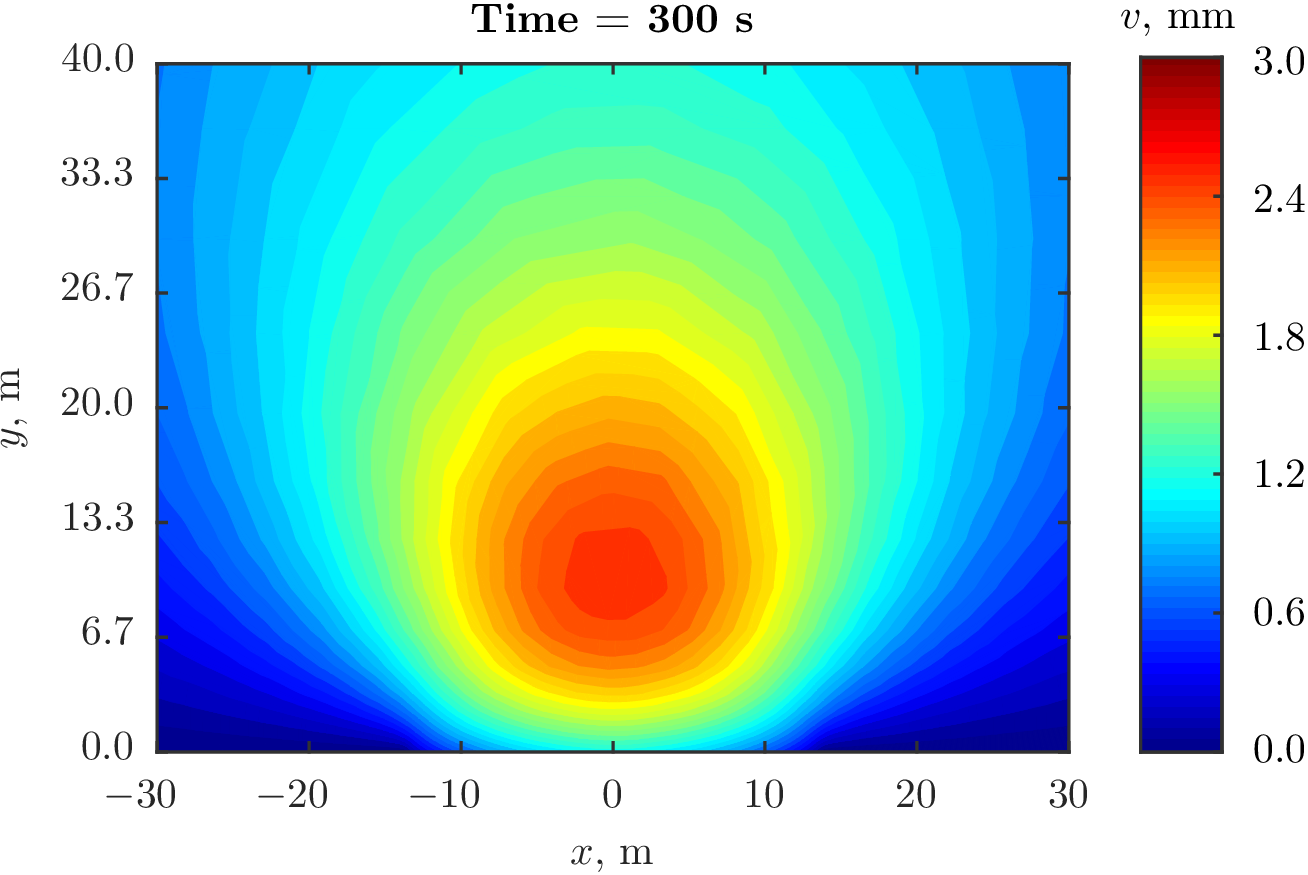}
		\centering {\bf(a)}
		
	\end{minipage}
	\hspace{0.5cm}
	\begin{minipage}[t]{0.5\linewidth}
		
		\includegraphics[width=1\linewidth]{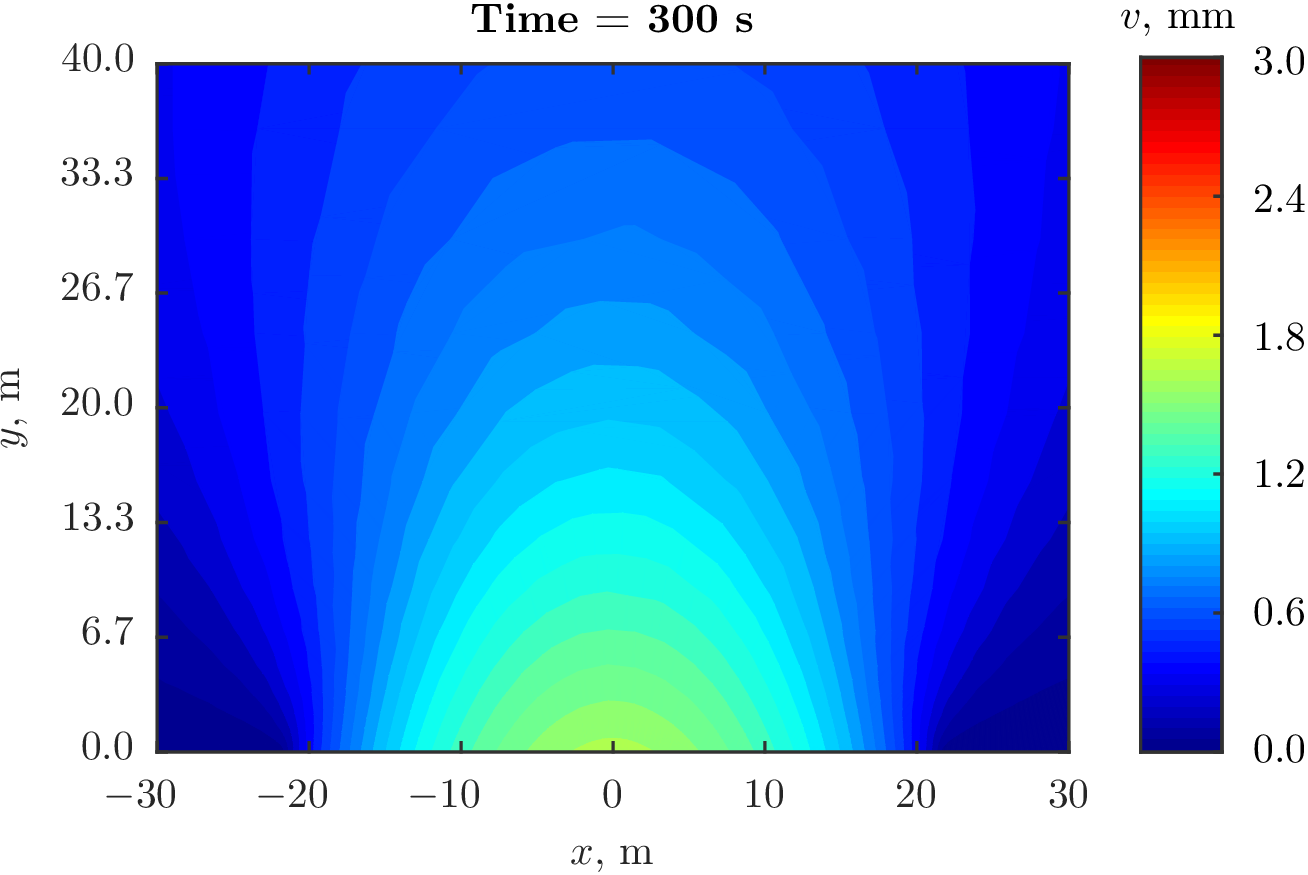}
		\centering{\bf(b)}
		
	\end{minipage}
	
	\caption{Vertical component $v$ of the displacement in (a) fully coupled case A, and (b) uncoupled case D at $t=300$ s. The physical parameters are $k_r = 10^{-14}$ m$^2$, ${S_\varepsilon=1.46\times 10^{-11}}$~Pa$^{-1}$}
	\label{fig:V3D11And00AlphaHiPermLoSeTimestep200}
	
\end{figure}

The second factor is the action of the rock deformation to the filtration. One could expect that, according to the First thumb rule, the leakoff of fluid from the fracture to the reservoir would be reduced due to the rock displacement as it happened in case C (see Section \ref{FirstThumbRule}). This effect is really taken place, although not in the fracture direction but towards the area of the maximal rate of rock displacement. Indeed, the pore pressure produces an additional stiffness of the rock near the fracture (the Second thumb rule) that decreases along $y$ coordinate. Therefore, the maximum value of the displacement is reached not on the fracture's wall, but at some distance from the fracture. This effect is demonstrated in Figure~\ref{fig:V3D11And00AlphaHiPermLoSeTimestep200}  where we compare the vertical displacement $v$ for cases of fully coupled case A (left) and uncoupled case D (right). According to the First thumb rule, fluid is attracted to the area of the highest rate of rock displacement, which is located at some distance from the fracture, and causes an extra leakoff from the fracture.

\begin{figure}[!htb]
		\begin{minipage}[t]{0.5\linewidth} 
			
			\includegraphics[width=1\linewidth]{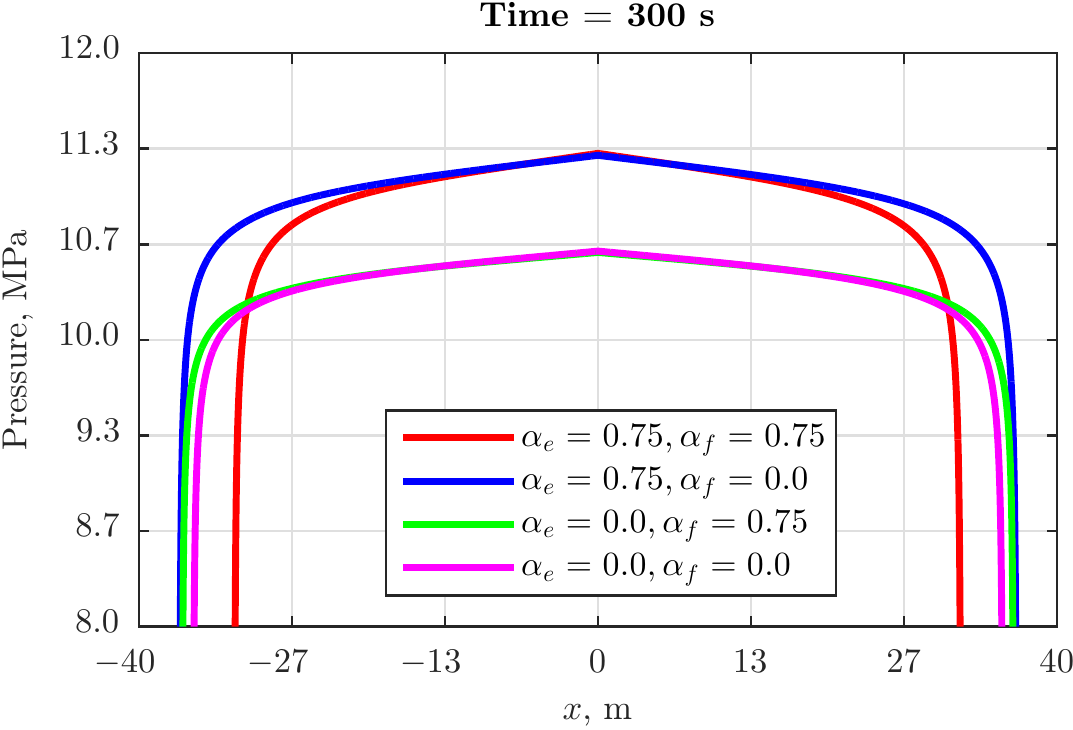}
			\centering {\bf(a-1)}
			
		\end{minipage}
		\hspace{0.5cm}
		\begin{minipage}[t]{0.5\linewidth}
			
			\includegraphics[width=1\linewidth]{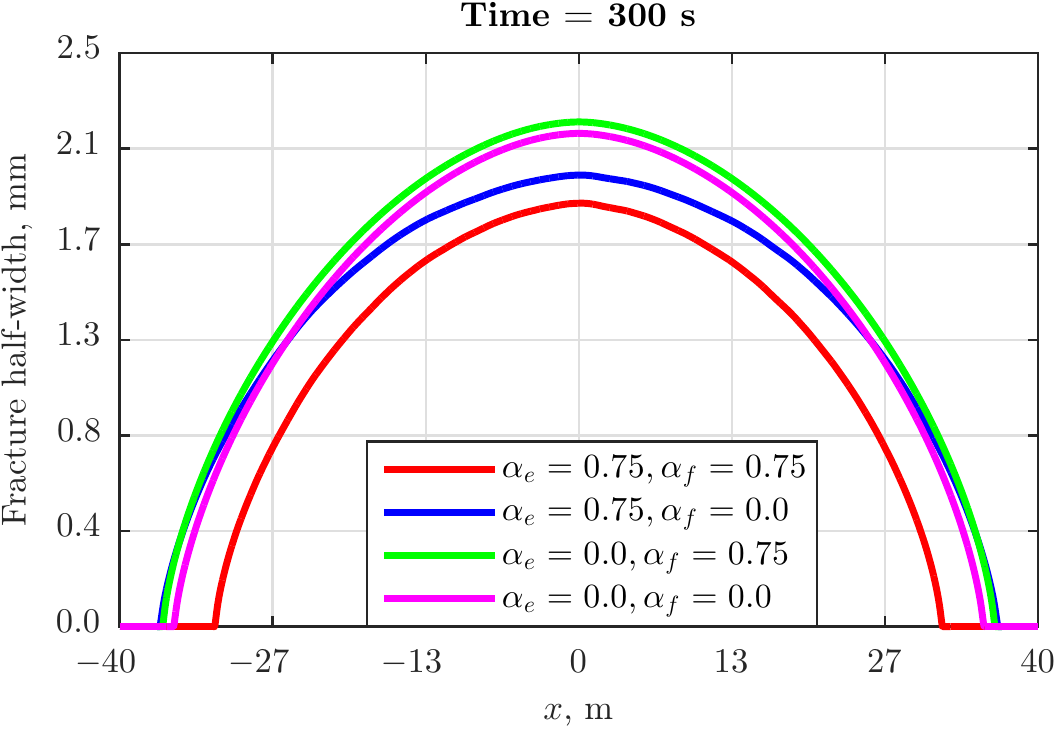}
			\centering{\bf(b-1)}
			
		\end{minipage}
		\smallskip
		

		\begin{minipage}[t]{0.5\linewidth} 
			
			\includegraphics[width=1\linewidth]{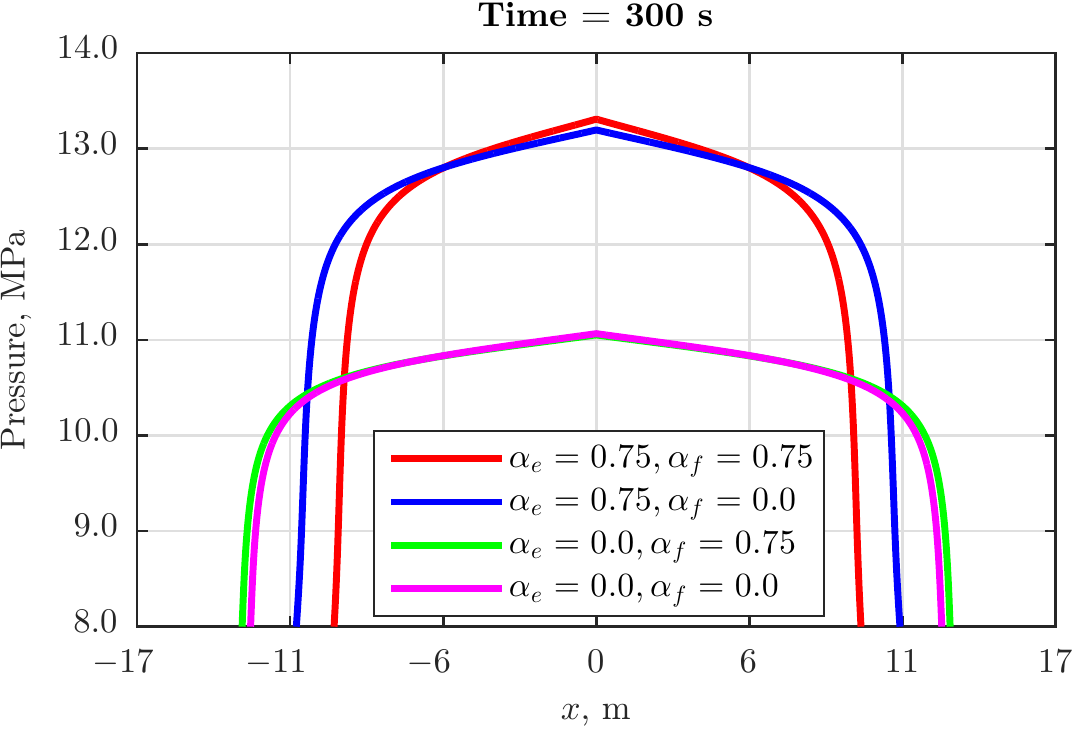}
			\centering {\bf(a-2)}
			
		\end{minipage}
		\hspace{0.5cm}
		\begin{minipage}[t]{0.5\linewidth}
			
			\includegraphics[width=1\linewidth]{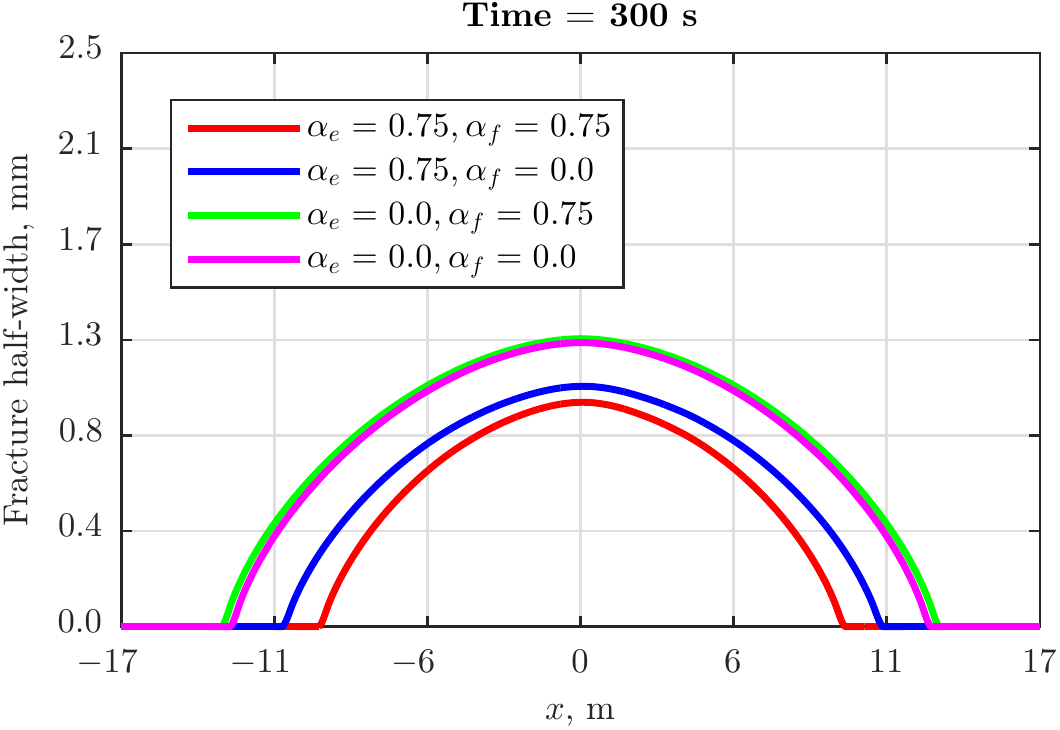}
			\centering{\bf(b-2)}
			
		\end{minipage}
		\smallskip		
		
		
		\begin{minipage}[t]{0.5\linewidth} 
			
			\includegraphics[width=1\linewidth]{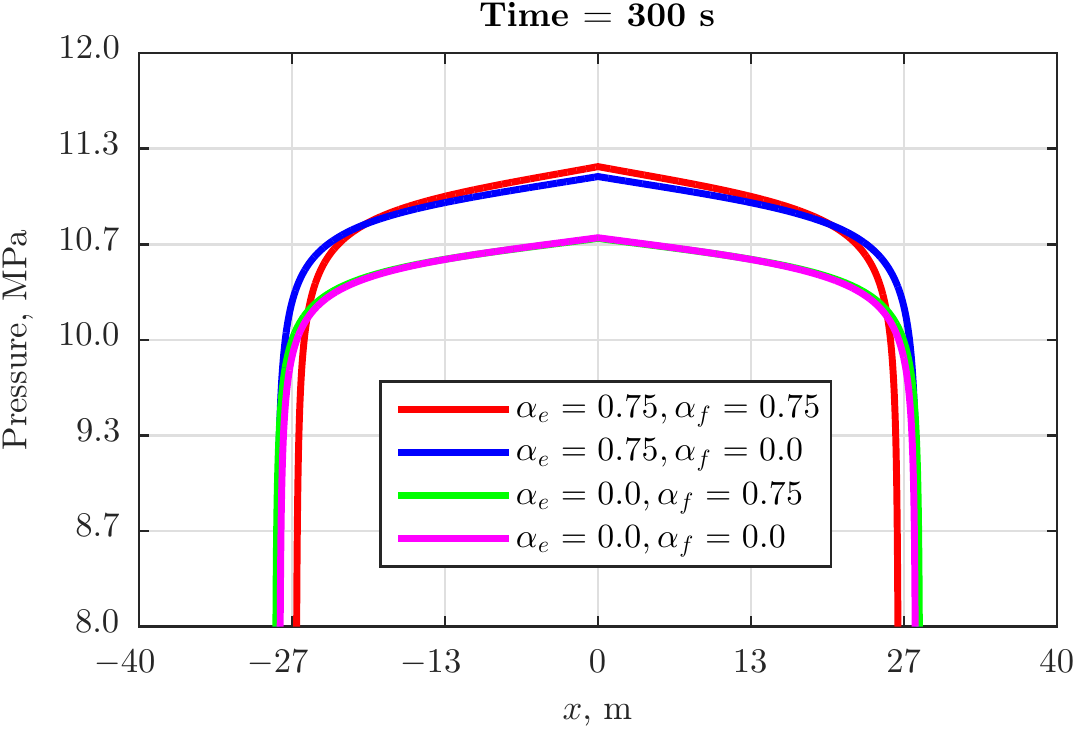}
			\centering {\bf(a-3)}
			
		\end{minipage}
		\hspace{0.5cm}
		\begin{minipage}[t]{0.5\linewidth}
			
			\includegraphics[width=1\linewidth]{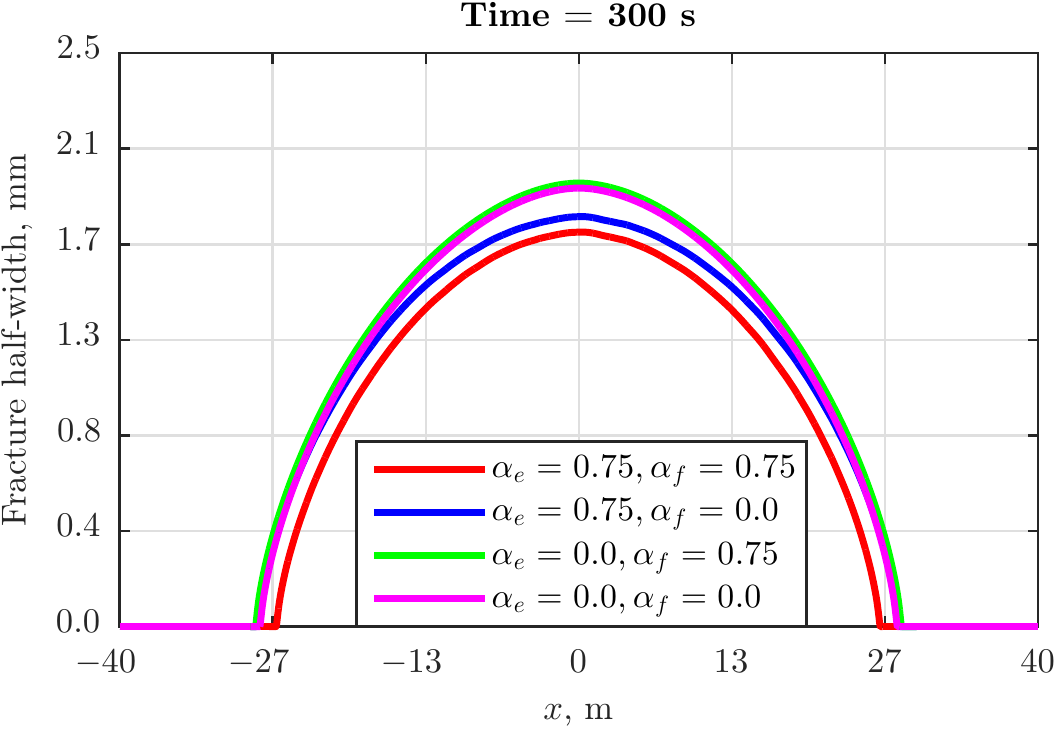}
			\centering{\bf(b-3)}
			
		\end{minipage}	
		\caption{Pressure (a) and fracture half-width (b) along the fracture for coupled, uncoupled and partially coupled cases at $t=300$ s. The physical parameters are 
		1) low permeability, low storativity: $k_r = 10^{-15}$ m$^2$, $S_\varepsilon = 1.46\times 10^{-11}$~Pa$^{-1}$,
		2) high permeability, high storativity: $k_r = 10^{-14}$ m$^2$, $S_\varepsilon = 7.3\times 10^{-11}$~Pa$^{-1}$,
		3) low permeability, high storativity: $k_r = 10^{-15}$~m$^2$, $S_\varepsilon =7.3\times 10^{-11}$~Pa$^{-1}$.}
		\label{fig:pressureAndWidthProfileAlphaZeroNonZeroLoPermHiSeTimestep200}		
	\end{figure}

As a conclusion to this section we note that for the relatively high rock permeability $k_r = 10^{-14}$ m$^2$, the pore pressure plays a significant role in the re-distribution of stresses near the hydraulic fracture and causes the change of about 20\% in the fracture geometric parameters. We also note that for the lower rock permeability or the higher storativity, all the mentioned tendencies are preserved but appear in less extent as shown in Figure~
\ref{fig:pressureAndWidthProfileAlphaZeroNonZeroLoPermHiSeTimestep200}.

\section{Conclusion and discussion}
In this work we presented a numerical model for the growth of a hydraulic fracture in a poroelastic reservoir. The model is based on Biot poroelasticity equation where the mutual influence of the elastic stress and the pore pressure is taken into account. The hydraulic fracture is driven by the fluid that is injected into the fracture at a fixed point. As the pressure of the injected fluid overcomes the confining stress and the rock toughness, the fracture propagates along a prescribed straight path perpendicular to the confining stress such that left and right fracture's tips propagates independently. Flow of fracturing fluid within the fracture is governed by the lubrication equation (i.e., mass conservation joined with the Poiseuille law for the flow rate). The fluid exchange between the fracture and the porous reservoir is described by the source/sink term in the lubrication equation proportional to the normal derivative of the pore pressure over the fracture's wall. 

The rock failure is treated by the cohesive zone approach where the presence of cohesive forces inverse proportional to the fracture's opening in the vicinity of fracture's tip is supposed. The integral value of the cohesive forces is selected according to the prescribed rock toughness. The problem is brought to the weak form and solved via the finite element method. In the numerical algorithm we do not distinguish the fracture's tip neither do we rebuild the numerical mesh according to the fracture propagation. Instead, we assume that the fracture is propagating along the known straight path and use the penalty method in order to guarantee the non-overlapping of the opposite fracture's walls. The nonlinearity of the lubrication equation and the geometric nonlinearity related to the unknown positions of fracture's tips is resolved by iterations by Newton-Raphson method. The advantage of our approach is that we do not need to reconstruct the computational mesh according to the tips propagation. 

The correctness of the method is verified by numerical computation of the order of convergence of the method as well as by the comparison with known analytical solutions and numerical solutions. In particular, we compare our computations with the results of paper \cite{Carrier_Granet} for different flow regimes: storage-toughness, leakoff-toughness, storage-viscosity, and leakoff-viscosity dominated regimes and demonstrated good agreement of solutions. 

The constructed numerical algorithm for calculation of fracture propagation allows us to perform a thorough analysis of the influence of the pore pressure to the distribution of stresses near the fracture and to the geometric parameters of the fracture. We found that there is a strong mutual influence between the rate of the rock displacement and the filtration of the pore fluid. First, the rock deformation causes a filtration in the direction opposite to the rate of the material displacement. Second, the pore fluid creates an additional hydrostatic pressure that effectively stiffens the rock and create the so-called backstress near fracture's walls. Combined, these two factors provide a significant change of fracture's length up to 20\% for the rock permeability of  $k_r = 10^{-14}$ m$^2$. For the lower rock permeability the mentioned effects are dumped. The influence of the pore pressure to the fracture propagation is greater for higher Biot's number $\alpha$. The shortest fracture is obtained for the highest $\alpha$. The overall conclusion is that the correct account for the fluid exchange between the fracture and the reservoir requires not only computation of the fluid leakoff according to Carter's or any similar formula, but also consideration of the redistribution of stresses near the fracture and their back influence to the leakoff rate.

In the forthcoming paper we plan to investigate the influence of the pore pressure and inhomogeneity of rock physical parameters (permeability, confining stresses, etc.) to the symmetry and dynamics of fracture propagation.

\section*{Acknowledgements}
The work is supported by RFBR (grant 16-01-00610) and President grant for support of Leading scientific schools (grant NS-8146.2016.1).



\end{document}